\documentclass[%
superscriptaddress,
secnumarabic,
twocolumn,
amssymb,
amsmath,
nobibnotes,
aps,
prd,
showkeys,
nofootinbib
]{revtex4-1}
\usepackage[T1]{fontenc}
\usepackage{tgtermes}
\usepackage{amsmath}
\usepackage[Symbol]{upgreek}

\newcommand{\di}{\mathrm{d}}
\newcommand{\csq}{\cos^2\!}
\newcommand{\ssq}{\sin^2\!}
\newcommand{\vph}{\varphi}
\newcommand{\vth}{\vartheta}
\newcommand{\cala}{\mathcal{A}}
\newcommand{\calb}{\mathcal{B}}

\usepackage{graphicx}
\usepackage{dcolumn}
\usepackage{bm}
\usepackage{xcolor}
\usepackage[utf8]{inputenc}
\definecolor{navy}{RGB}{0,0,150}
\usepackage[colorlinks=true,allcolors=navy]{hyperref}
\usepackage{fancyhdr}
\begin{document}

\preprint{APS/123-QED}

\title{Hamilton-Jacobi equation for spinning particles near black holes}

\author{Vojt{\v e}ch Witzany}
\email{vojtech.witzany@asu.cas.cz}
\affiliation{%
Astronomical  Institute  of  the  Academy  of  Sciences  of  the  Czech  Republic,
Bo\v{c}n{\'i}  II  1401/1a,  CZ-141  00  Prague,  Czech  Republic
}%
\affiliation{%
Center of Applied Space Technology and Microgravity (ZARM), 
Universit{\"a}t Bremen, 
Am Fallturm 2,  
D-28359 Bremen, Germany
}%

\date{\today}

\begin{abstract}
A compact stellar-mass object inspiralling onto a massive black hole deviates from geodesic motion due to radiation-reaction forces as well as finite-size effects. Such post-geodesic deviations need to be included with sufficient precision into wave-form models for the upcoming space-based gravitational-wave detector LISA. I present the formulation and solution of the Hamilton-Jacobi equation of geodesics near Kerr black holes perturbed by the so-called spin-curvature coupling, the leading order finite-size effect. In return, this solution allows to compute a number of observables such as the turning points of the orbits as well as the fundamental frequencies of motion. This result provides one of the necessary ingredients for waveform models for LISA and an important contribution useful for the relativistic two-body problem in general.
\end{abstract}

\maketitle
\newpage


\tableofcontents

\section{\label{sec:intro}Introduction}


The interaction of an astrophysical object with a background gravitational field is characterized by its physical size and gravitational radius. If both of these are much smaller than the curvature radius (variability length) of the background, the center of mass of the body will follow an almost geodesic trajectory in the surrounding space-time \citep{poisson2011,harte2012,barack2018}. The post-geodesic corrections to the motion will then scale with powers of the ratio of either the physical size or gravitational radius of the body with respect to the curvature radius. Technically, the whole object is understood in this approximation as a ``particle'' carrying mass multipole moments, and the corrections are expressed as radiation-reaction forces as well as interactions of the multipoles of the body with the background field. This post-geodesic approach is also often called the ``self-force program'', referring to the fact that the additional forces appear due to the object-specific interaction with the background rather than only due to the background field itself.

The mentioned post-geodesic expansion is well suited to describe gravitational-wave inspirals of stellar-mass compact objects into massive black holes, which are one of the key sources of gravitational radiation for the upcoming space-based detector LISA \citep{gair2009,amaro2017}. For these so-called extreme mass ratio inspirals (EMRIs), both of the aforementioned expansion parameters become proportional to the ratio $q \equiv \mu/M \sim 10^{-4}-10^{-11}$, where $M, \mu$ are the masses of the primary and secondary of the binary respectively.

{\em EMRI phasing.} Assuming that a geodesic in the field of the primary is fully determined by some set of constants of motion (orbital parameters) and phase variables, we can sort the post-geodesic deviations into two classes based on their long-term effects \citep{hinderer2008}. First, the gravitational radiation will carry away a part of the constants of motion such as energy and angular momentum, and will thus lead to a long-term change in their values. These effects are conventionally called {\em dissipative}. 

Second, the deviations will cause the constants of motion to oscillate as well as to change the rates at which the trajectory goes through its phases. In the long term, this can be characterized as a change of the frequency with which we see the orbit is passing through its phases or, alternatively, as a secular accumulation of a set of post-geodesic phase-shifts. This second type of effects is usually called {\em conservative}.\footnote{Note that the ``dissipative'' effects do not correspond to any transfer of information from the orbit into microscopic degrees of freedom, they correspond to the transfer of information into macroscopic gravitational-wave degrees of freedom. On the other hand, the ``conservative'' effects cannot be completely captured as a Lagrangian or Hamiltonian perturbation to the original Lagrangian or Hamiltonian, so the dissipative/conservative nomenclature should be understood exactly as stated in the text.}

It is now generally accepted that wave-form models that will allow LISA to accurately estimate the parameters of the sources have to include all dissipative post-geodesic corrections to the equations of motion up to $\mathcal{O}(q^2)$ and conservative corrections up to $\mathcal{O}(q)$ \citep{hinderer2008,poisson2011,barack2018}. The only $\mathcal{O}(q)$ correction to the equations of motion due to the finite size of the body is the so-called spin-curvature coupling and it is entirely conservative. 

{\em Spin-curvature coupling.} The spin-curvature coupling arises due to the fact that different parts of a relativistically rotating body interact with the background space-time in a way that ends up exerting a ``spin force'' on the center of mass that is proportional to the angular momentum of the body about its center and the background curvature. The dissipative $\mathcal{O}(q^2)$  effects due to the finite size of the body then currently seem to be only cross-terms of radiation reaction and spin; the gravitational radiation will gradually carry away a part of the internal angular momentum of the body or irreversibly ``steer'' its direction, and the fluctuations to the orbit due to the spin force will slightly modulate the power with which the geodesic constants of motion are radiated away.

There is a long history of works studying the motion of spinning test bodies in black hole space-times, most of which was reviewed in Refs. \citep{semerak1999,kyrsem}. A subset of these studies then used the spinning test body motion to generate and study outgoing gravitational waves \citep{tanaka1996,suzuki1997,tominaga2001,Harms2016,lukes2017}. Yet another thread of research computed the precession of the spin of the particle as a gauge invariant probe of the self-force, which is convenient for comparison with other approaches to the relativistic two-body problem \citep{dolan2014,akcay2017,kavanagh2017,akcay2017b,bini2018}. Various perturbative formalisms for the computation of post-geodesic corrections to orbital motion in black hole space-times due to the spin-curvature coupling were previously formulated \citep{mashhoon2006,singh2008,bini2011,bini2011a}, typically focusing on numerical solutions of the equations or on special classes of orbits. Finally, concrete computations of EMRIs with spin effects were carried out in Refs. \citep{huerta2011,huerta2012,burko2015,Warburton2017,will2017}.

{\em Resonances.} It is routinely observed that perturbation-theory techniques fail in the neighborhood of orbits for which unperturbed fundamental frequencies reach integer ratios \citep{arnold2007}. Generally, the perturbation of order $\epsilon$ then causes a qualitative change in the behaviour of an $\mathcal{O}(\sqrt{\epsilon})$ volume of orbits in the neighborhood of these so-called resonant orbits. This issue has also been identified in the case of perturbative expansion of EMRIs \citep{flanagan2012}, the locations of the {\em orbital} resonances were computed by \citet{brink2015,brink2015b}, the physical consequences of the passage of an EMRI through a resonance were studied in Refs. \citep{flanagan2014,vandemeent2014,vandemeent2014b,lewis2017}, and consequences of resonant effects for LISA science were investigated in Refs. \citep{apostolatos2009,lukes2010,ruangsri2014,berry2016}. For the case of the spin perturbation the spin force could cause resonances of width $\mathcal{O}(\sqrt{q})$ and one can also have {\em spin-orbital} resonances corresponding to an integer ratio of frequency of the evolution of the spin and some of the orbital frequencies. 

{\em Hidden symmetry.} The Kerr black hole has a ``hidden symmetry'' (see section \ref{subsec:kerrmet}) and it was often investigated whether this implies new conserved quantities along the motion of spinning particles. \citet{ruediger1,ruediger2} found approximate integrals of motion for spinning particles under the Tulczyjew-Dixon supplementary spin condition in general space-times with hidden symmetry (see section \ref{subsec:constint}), and \citet{apostolatos1996} and \citet{kunst2016} found such integrals of motion in Schwarzschild space-time under the Mathisson-Pirani and Newton-Wigner conditions respectively. Conserved quantities due to the hidden symmetry for the motion of \textit{semi-classical} spinning particles (particles with supersymmetry on the wordline) were then studied in Refs. \citep{gibbons1993,tanimoto1995,ahmedov2009,kubiznak2012}. Additionally, it was observed in the frequency-domain analysis by \citet{ruangsri2016} that the hidden symmetry seems to ``protect'' the spin-perturbed orbits from resonances and chaos.

The purpose of this paper is to present a complete scheme for the computation of the conservative spin-curvature corrections to geodesic motion. The core of the computational scheme is an analytical perturbative solution of the Hamilton-Jacobi equation for the particle with spin orbiting a Kerr black hole on a generally inclined and eccentric trajectory.

The solution to the Hamilton-Jacobi equation is almost, but not quite separable in Boyer-Lindquist coordinates and it leads to two separation constants equivalent to those of \citet{ruediger1,ruediger2}. Consequently, the order (or number) of the equations of motion is reduced to half while not yielding them fully separable. Nevertheless, this still allows to analytically solve for the turning points, and to obtain actions and shifts of the fundamental frequencies of motion in terms of simple quadratures. These results resolve the question of the conservative spin-induced post-geodesic phase shifts in EMRIs (appearing at $\mathcal{O}(q)$ in the equations of motion) and provide interesting prospects for the computation of the dissipative effects (appearing at $\mathcal{O}(q^2)$ in the equations of motion). 

The paper is organized as follows. I start with stating all the important relations and definitions for time-like geodesics in Kerr space-time in Section \ref{sec:kerrgeo}, and I introduce the Hamiltonian formalism for spinning particles along with the set of coordinates and the adapted tetrad I use in Section \ref{sec:hamspin}. Then I present the perturbative solution to the respective Hamilton-Jacobi equation along with a discussion of separation constants in Section \ref{sec:hamjac}. Finally, the computation of various properties of the spin-perturbed orbits such as turning points or frequencies of motion are discussed in Section \ref{sec:orb}. Details of the derivations of Section \ref{sec:orb} are discussed in the Appendix.

I use the $G=c=1$ geometrized units and the (-+++) signature of the metric. The ordinary derivative with respect to $x^\mu$ is denoted by an index $\mu$ preceded by a comma, and a covariant derivative with an index preceded by a semi-colon. My convention for the Riemann tensor $R^{\mu}_{\;\nu\alpha\beta}$ is such that $a_{\mu;\alpha\beta} - a_{\mu;\beta\alpha} = R^{\nu}_{\;\mu\alpha\beta} a_{\nu}$ for a generic $a_\mu$. $\eta^{\mu\nu}$ with any indices is the Minkowski tensor, and ${\delta^\mu}_\nu$ denotes the Kronecker delta.


\section{Kerr geodesics} \label{sec:kerrgeo}

Geodesics in a given space-time are perhaps the most faithful conveyors of its geometry. Kerr geodesics and their properties will be important in various different ways in the upcoming sections; they will generate the tetrad used in section \ref{sec:hamspin}, and serve as a ``zeroth-order'' unperturbed or fiducial system in sections \ref{sec:hamjac} to \ref{sec:orb}. Hence, I will now briefly summarize the necessary notation and details about them.

\subsection{Kerr metric} \label{subsec:kerrmet}
The nonzero components of the inverse Kerr metric in Boyer-Lindquist coordinates $t,\vph,r,\vth$ read \citep{boyer1967}
\begin{align}
\begin{split}
  & g^{tt} = -\frac{\Sigma (r^2 + a^2) + 2 M r a^2 \ssq \vth}{\Delta \Sigma}\,,\\
  & g^{t\vph} =g^{\vph t} = -\frac{-2 M ra}{\Delta \Sigma }\,,\\
  & g^{\vph \vph}= \frac{\Sigma - 2 M r}{\Delta \Sigma \ssq \vth}\,,\\
  & g^{rr} = \frac{\Delta}{\Sigma}\,,\\
  & g^{\vth\vth} = \frac{1}{\Sigma}\,,\\
  \label{eq:kerrinv}
  \end{split}
\end{align}
where $\Sigma = r^2 + a^2 \csq \vth$ and $\Delta = r^2 - 2M r + a^2$. The Kerr metric is independent of $t,\varphi$, which makes specific energy $E \equiv - u_t$ and specific azimuthal angular momentum $L \equiv u_\varphi$ constants of geodesic motion, where $u^\mu = \di x^\mu/\di \tau, u^\mu u_\mu = -1$ is the four-velocity. 

The Kerr metric possesses a Killing-Yano tensor $Y_{\mu\nu}=-Y_{\nu\mu},\, Y_{\mu\nu;\kappa} = -Y_{\mu\kappa;\nu} $with the components \citep{floyd1973,penrose1973}
\begin{align}
\begin{split}
& Y_{rt} = -Y_{tr} =   a \cos \vth \,,\\
& Y_{r\varphi} = -Y_{\varphi r} =  -a^2\cos \vth \sin^2 \! \vth \,,\\
& Y_{\vartheta \varphi} = -Y_{\varphi \vartheta} = (r^2 + a^2) r \sin \vth   \,,\\
&  Y_{\vartheta t} = -Y_{t\vartheta} =  -a r \sin \vth \,,\\
& Y_{t\varphi} = - Y_{\varphi t} = Y_{r\vartheta} = - Y_{\vartheta r} = 0 \,. \label{eq:KYcomponents}
\end{split}
\end{align}
The properties of the Killing-Yano tensor imply parallel transport of a vector $l^\mu = Y^{\mu\nu} u_\nu$ along geodesics, $ l^\mu_{\;\;;\nu}u^\nu= 0$. This also implies the conservation of the square of this vector $\di K/\di \tau = 0$, $K \equiv l^\mu l_\mu$. It is natural to interpret $l^\mu$ as some sort of specific angular momentum vector, and $K$, also known as the Carter constant \citep{carter1968}, as specific angular momentum squared. The existence of the tensor $Y_{\mu\nu}$ in the Kerr metric is considered to be a ``hidden symmetry'' of the space-time.

\subsection{Four-velocities}
The Hamilton-Jacobi equation for geodesics in Kerr space-time is separable in Boyer-Lindquist coordinates, which was shown by \citet{carter1968}. Additionally, if one uses the Mino time $\di \lambda / \di \tau = 1/\Sigma$ \citep{carter1968,mino2003}, the equations of motion in the $r,\vth$ plane also completely decouple and we obtain
\begin{subequations}
\begin{align}
    &\frac{\di r}{\di \lambda} = \pm\sqrt{R(r)}\,,\\
    &\frac{\di \vth}{\di \lambda} = \pm\frac{\sqrt{\Theta(\vth)}}{\sin \vth}\,,\\
    & R(r) = -(K+r^2)\Delta + \left(E(r^2 +a^2) - a L \right)^2 \,,\\
    & \Theta(\vth) = (K - a^2 \csq \vth)\ssq\vth - \left(L -a E \ssq \vth \right)^2\,,
\end{align}
\end{subequations}
where the factor $1/\sin \vartheta$ in the $\vth$ equation comes from the fact that it is often practical to use the variable $\zeta =\cos \vth$ in the integration as well as symbolic manipulation of the equations. The $\varphi,t$ motion can be integrated once $r(\lambda),\vth(\lambda)$ is known.
Various other formulae and results for the geodesic motion in Kerr space-time were summarized by \citet{chandrasekharBH}. 

\subsection{Characterization by roots}

We saw in the last section that a geodesic in Kerr space-time is uniquely specified by the set of constants of motion $K,E,L$. Nevertheless, it is sometimes useful to instead specify the geodesics by their turning points. The functions $R,\Theta$ can be rewritten as
\begin{subequations}
\begin{align}
    & R(r) = (1 - E^2)(r_{1\rm g} -r)(r- r_{2\rm g})(r- r_{3\rm g})(r- r_{4\rm g})\,,\\
    & \Theta(\vth) = a^2(1 - E^2)(z_{+\rm g} - \csq\vth)(z_{-\rm g} - \csq\vth)\,,
\end{align}
\end{subequations}
where $r_{1\rm g},r_{2\rm g},r_{3\rm g},r_{4\rm g}$ are the roots of the function $R(r)$ ordered by magnitude from largest to smallest, and $z_{\pm \rm g}$ are the $\csq \vth$ roots of $\Theta$. The geodesic itself will oscillate in the ``box'' $\cos \vth \in (-\sqrt{z_{-\rm g}},\sqrt{z_{- \rm g}})\,,r\in (r_{2 \rm g},r_{1 \rm g})$. To obtain an intuitive picture of the orbit, it is useful to parametrize the radial turning points by eccentricity $e$ and semi-latus rectum $p$ \citep{schmidt2002}
\begin{align}
    r_{1 \rm g} = \frac{p}{1-e}\,,\; r_{2 \rm g} = \frac{p}{1 + e}\,.
\end{align}
The set of orbital parameters $p,e,\sqrt{z_{-\rm g}}$ specify a geodesic uniquely, and the relation to the $K,E,L$ specification was given by \citet{drasco2004}. Furthermore, \citet{fujita2009} gave analytical expressions for fundamental frequencies of motion based on this formalism (for similar analytical results in space-times generalizing the Kerr space-time, see Refs. \citep{hackmann2010,kagramanova2010,hackmann2012}). Many of the aforementioned formulae are implemented in the KerrGeodesics Mathematica package \citep{kerrgeo}.


\section{Hamiltonian formalism for spinning particles} \label{sec:hamspin}
The motion of a spinning body expanded to pole-dipole order is characterized by the position of its center of mass $x^\mu$, its momentum (stress-energy monopole) $P^\mu$, and internal angular momentum (stress-energy dipole) $S^{\mu\nu} = -S^{\nu\mu}$. Any body of finite size will, in fact, have an infinite tower of multipoles, but I neglect the influence of quadrupole and higher order moments since I am concerned here only with the leading-order finite-size effects on the orbital motion.

The Mathisson-Papapetrou-Dixon (MPD) \citep{mathisson1937,papapetrou1951,dixon1964} equations that govern the evolution of the body to pole-dipole order then read \citep{mathisson1937,papapetrou1951,dixon1964}
\begin{subequations}
\label{eq:mpd}
\begin{align}
&\frac{\mathrm{D}{P}^\mu}{\di \lambda} = -\frac{1}{2} R^{\mu}_{\; \nu \kappa \lambda} \frac{\di x^\nu}{\di \lambda} S^{\kappa \lambda} \,,\\
&\frac{\mathrm{D}{S}^{\kappa \lambda}}{\di \lambda} = P^\kappa \frac{\di x^\lambda}{\di \lambda} - P^\lambda \frac{\di x^\kappa}{\di \lambda}\,,
\end{align}
\end{subequations}
where $\mathrm{D}/\di \lambda$ is a covariant derivative with respect some parameter $\lambda$ along the trajectory $x^\mu(\lambda)$ (not necessarily the proper time).

The MPD equations require a specification in which frame $V^\mu$ the center of mass as well as the multipoles are computed. Consequently, the electric-type dipole $S^{\mu\nu}V_\nu$ vanishes in this frame. I choose this frame as parallel to $P^\mu$, $S^{\mu\nu}P_\nu = 0$, which is known as the Tulczyjew-Dixon or ``covariant'' supplementary spin condition \citep{tulczyjew1959,dixon1970} (see \citep{costa2015} for a review of other choices of $V^\mu$). A concise summary of the MPD equations under the Tulczyjew-Dixon condition in Kerr space-time is given in Ref. \cite{semerak1999}.

I now briefly introduce the Hamiltonian formalism for the MPD equations, since the knowledge of the Hamiltonian and canonical coordinates covering the phase space are prerequisites for the formulation of the Hamilton-Jacobi equation in section \ref{sec:hamjac}. An important point is the introduction of a specifically oriented set of coordinates through an adapted tetrad in section \ref{subsec:marck}, the choice of which ultimately allows for the partial separation of the Hamilton-Jacobi equation.

\subsection{Hamiltonian for Tulczyjew-Dixon condition} \label{subsec:ham}

Under the Tulczyjew-Dixon condition $S^{\mu\nu}P_\nu = 0$ the Hamiltonian for the motion of the spinning body is given as \citep{spinpap}
\begin{subequations} \label{eq:tdhamspec}
\begin{align}
& H_\mathrm{TD} = \frac{1}{2}  \left( g^{\mu\nu} - \gamma^{\mu\nu}\right) U_\mu U_\nu \cong -1 \,, 
\\
& \gamma^{\mu\nu} \equiv \frac{4 s^{\nu\gamma} R^\mu_{\; \gamma \kappa \lambda} s^{\kappa \lambda} }{4 +  R_{\chi \eta \omega \xi} s^{\chi \eta} s^{\omega \xi}}  \,,
\end{align}
\end{subequations}
where the variables are specific momenta defined as $U^\mu \equiv P^\mu/\mathcal{M},\,s^{\mu\nu} \equiv S^{\mu\nu}/\mathcal{M}\,,\mathcal{M}^2 = - P^\mu P_\mu$. The Hamiltonian generates the MPD equations \eqref{eq:mpd} when used along the Poisson bracket
\begin{subequations} \label{eq:poiss}
\begin{align}
\{x^\mu,x^\nu\} &= 0\,,\\
 \{x^\mu, U_\nu\} &=  \delta^\mu_\nu \label{eq:poissxpcov} \,, \\
\{U_\mu, U_\nu\} & = -\frac{1}{2} R_{\mu\nu \kappa \lambda} s^{\kappa \lambda} \label{eq:poisspp} \,, \\
 \{s^{\mu\nu}, U_\kappa\} & = -\Gamma^\mu_{\;\lambda \kappa} s^{\lambda \nu} - \Gamma^\nu_{\; \lambda \kappa} s^{\mu \lambda} \label{eq:poissSp} \,, \\
 \{s^{\mu\nu}, x^\kappa\} & = 0 \label{eq:poissSx} \,, \\
\begin{split}
 \{s^{\mu\nu}, s^{\kappa \lambda}\} & =  g^{\mu \kappa}s^{\nu \lambda} - g^{\mu \lambda}s^{\nu \kappa} + g^{\nu \lambda}s^{\mu \kappa} \\ & - g^{\nu \kappa}s^{\mu \lambda} \label{eq:poissSS} \,,
\end{split}
\end{align}
\end{subequations}
where the parameter $\lambda$ of the evolution is close to proper time, $\di \lambda = \di \tau + \mathcal{O}(s^2)$, and such that $U_\mu (\di x^\mu/\di \lambda) = -1$ \cite{dixon1970,ehlers1977}.

The equality $\cong$ in \eqref{eq:tdhamspec} is fulfilled under the condition that the  initial data is chosen such that $U^\mu U_\mu = -1$ and $s^{\mu\nu} U_\nu = 0$, and these relations are then also conserved along the motion. An additional quantity that is conserved along the motion is the magnitude of the specific angular momentum $s = \sqrt{s^{\mu\nu}s_{\mu\nu}/2}$. For more details on the Hamiltonian formalism for spinning particles see Ref. \cite{spinpap}.

\subsection{Canonical coordinates}
Consider a tetrad basis $e_{C\mu}, C=0,1,2,3, e_{C \mu} e_{D}^\mu = \eta_{CD}$ and variables 
\begin{subequations}
\begin{align}
&\mathcal{U}_\mu \equiv U_\mu +\frac{1}{2}e_{C\nu;\mu}e^\nu_D s^{CD} \,,\\
&s_{CD} \equiv s^{\mu\nu}e_{C\mu}e_{D\nu}
\end{align}
\end{subequations}
Now the variables $x^\mu,\mathcal{U}_\mu$ are canonically conjugate, $\{x^\mu,\mathcal{U}_\nu\} = \delta^\mu_\nu$ and zero for any other bracket involving the variables. The spin sector is covered by two canonically conjugate pairs of coordinates and momenta $\phi,\cala$ and $\psi,\calb$, which parametrize $s^{CD} = - s^{DC}$ as \cite{spinpap}
\begin{subequations}
\label{eq:spparam}
\begin{align}
&s^{01} = \mathcal{C} \left[ \cala \cos(2 \phi - \psi) + (\cala + 2 \calb + 2s) \cos \psi \right] \,,\\
&s^{02} = \mathcal{C} \left[ \cala \sin(2 \phi - \psi) + (\cala + 2 \calb + 2s) \sin \psi \right] \,,\\
&s^{03} = - 2\mathcal{C}\mathcal{D}  \cos(\phi -\psi) \,,\\
&s^{12} = \cala+\calb+s\,,\\
&s^{23} = \mathcal{D} \cos\phi\,,\\
&s^{31} = \mathcal{D} \sin\phi\,,\\
& \mathcal{C} =  -\frac{\sqrt{\calb(\calb + 2s)}}{2(\calb+s)} \,, \\
& \mathcal{D} =  \sqrt{-\cala(\cala+2\calb + 2s)} \,. 
\end{align}
\end{subequations}
The physical ranges of these coordinates are $\cala\in[-2(\calb+s),0], \, \calb \in [0, \infty)$, and $\phi,\psi \in [0,2\uppi)$. The inverse transform is given in \citep{spinpap} and one can then verify by direct computation from the brackets \eqref{eq:poiss} that the coordinates fulfill $\{\phi,\mathcal{A}\}=\{\psi,\mathcal{B}\} = 1$ and zero otherwise.

\subsection{Adapted tetrad} \label{subsec:marck}
One can notice from the previous section that different choices of the tetrad $e_{C\mu}$ correspond to a different covering of the phase space of the spinning particle by canonical coordinates. It is well known that choosing the right set of coordinates is often crucial to the analytical solution of a problem. I now introduce a special ``geodesic-adapted'' tetrad that will provide a useful basis for the computations in the next parts of the paper.

We start with taking a geodesic congruence with constants of motion $K_{\rm c}, E_{\rm c}, L_{\rm c}$ as the zeroth leg
\begin{align}
& u_{r\rm c} = \pm \frac{\sqrt{R(r;K_{\rm c}, E_{\rm c}, L_{\rm c})}}{\Delta}\,,\\
& u_{\vth \rm c} = \pm \frac{\sqrt{\Theta(\vth;K_{\rm c}, E_{\rm c}, L_{\rm c})}}{\sin \vth}\,,\\
& u_{\vph \rm c} = L_{\rm c} \,,\\
& u_{t \rm c} = - E_{\rm c}\,.
\end{align}
In other words, $e_{0 \mu} = u_{\mu \rm c}$. Now another leg of the tetrad can be generated by the antisymmetric Killing-Yano tensor $e_{3\mu} = Y_{\mu\nu}u^\nu_{\rm c}/\sqrt{K_{\rm c}}$. The last two legs are also generated by the Killing-Yano tensor as
\begin{align}
e^1_\mu &= \frac{1}{N_{(1)}} \left(K_{\mu \nu} + K_\mathrm{c} g_{\mu \nu} \right) u^\nu_{\rm c} \,,\\
e^2_\mu &= \frac{1}{N_{(2)}} \left(K_{\mu \nu} - \frac{K_\mathrm{c}^{(2)}}{K_\mathrm{c}} g_{\mu \nu} \right)Y^{\nu}_{\;\kappa} u^\kappa_{\rm c} \,,\\
 K_{\mu\nu} &\equiv Y_{\mu}^{\;\kappa}Y_{\nu \kappa}\,,\\
N_{(1)}^2&\equiv K_\mathrm{c}^{(2)} + K_\mathrm{c}^2 = (K_\mathrm{c} - r^2)(K_\mathrm{c} - a^2 \csq\vth)\,,\\
\begin{split}
N_{(2)}^2 &\equiv K_\mathrm{c}^{(3)} - (K_\mathrm{c}^{(2)})^2/K_\mathrm{c} \\
& = \frac{r^2 a^2 \csq \vth (K_\mathrm{c} - r^2)(K_\mathrm{c} - a^2 \csq\vth)}{K_\mathrm{c}}\,,
\end{split}\\
 K_\mathrm{c}^{(2)} &\equiv K_{\mu \nu}K^{\nu}_{\; \kappa} u^\mu_{\rm c} u^\kappa_{\rm c} \,,\\
 K_\mathrm{c}^{(3)} &\equiv K_{\mu \nu}K^{\nu}_{\; \kappa}K^{\kappa}_{\; \gamma} u^\mu_{\rm c} u^\gamma_{\rm c} \,.
\end{align}
It is then easy to verify that the tetrad is orthonormal and normalized, $e_{A \mu} e_{B}^\mu = \eta_{AB}$. Note that apart from the parameters $K_{\rm c}, E_{\rm c}, L_{\rm c}$ the tetrad also needs to be specified by the choices of the sign of $e_{0r},e_{0\vth}$, and also that it is defined only within the turning points of the congruence. Finally, when we compare our tetrad with that of \citet{marck}, we see that they are identical, even though they have been arrived to by different procedures. 

Thanks to the construction of the tetrad we have $e^\kappa_{3;\mu} e_0^\mu = 0$ and the only nonzero projection of the zeroth leg into the spin connection components is $e_{1;\mu}^\kappa e_{2\kappa} e^{\mu}_0 = -e_{2;\mu}^\kappa e_{1\kappa} e^{\mu}_0$, which reads
\begin{align}
\begin{split}
    & e_{2;\mu}^\kappa e_{1\kappa} e^{\mu}_0 = 
    \\
    & \frac{\sqrt{K_\mathrm{c}}}{\Sigma} \Big( \frac{E_\mathrm{c} (r^2 + a^2) - a L_\mathrm{c}}{ r^2 + K_\mathrm{c}} + a \frac{L_\mathrm{c} - a E_\mathrm{c} \sin^2 \! \vartheta}{K_\mathrm{c} -  a^2 \cos^2 \! \vartheta} \Big)\,.
\end{split}
\end{align}
We will see that this separable form of the projections will be crucial in solving the perturbative Hamilton-Jacobi equation.


\section{Hamilton-Jacobi equation} \label{sec:hamjac}
The Hamilton-Jacobi equation is obtained by substituting canonical momenta in the Hamiltonian by gradients of the action $W(t,\vph,r,\vth,\phi,\psi)$ with respect to their conjugate coordinates. Specifically, we have $\mathcal{U}_\mu \to W_{,\mu},\, \cala\to W_{, \phi},\,\calb \to W_{,\psi}$. In the case $s=0$ ($\cala=0,\calb=0$) the Hamilton-Jacobi equation corresponding to the Hamiltonian \eqref{eq:tdhamspec} reads
\begin{align}
    g^{\mu\nu} W^{(0)}_{,\mu} W^{(0)}_{,\nu} = -1\,, \label{eq:nuljac}
\end{align}
which is the Hamilton-Jacobi equation for the geodesic with the well known solution by \citet{carter1968} (see Section \ref{sec:kerrgeo}). 

Now we want to perturb the zeroth-order solution by adding terms linear in spin to equation \eqref{eq:nuljac}. However, it turns out that when we use the adapted tetrad presented in the previous section, the size of the connection terms change as we approach the turning points of the background congruence. Thus, I construct the solution in two steps that correspond to regions with different magnitudes of the connection terms.
\subsection{Swing region solution}
Let us first assume that we are in the ``swing region'' of the tetrad, that is, far away from the turning points of the congruence $u_{\rm c}^\mu$. Formally the swing region is specified as the range of $r,\vartheta$ for which $|r-r_{1,2\rm c}| \gg s$ and $r(|\!\cos \vartheta| - \sqrt{z_{-\rm c}}) \gg s$, where $r_{1,2\rm c},\sqrt{z_{-\rm c}}$ correspond to the turning points of the background congruence. 

Now we are looking for a swing-region solution to the action $W^{\rm (1\rm sw)}_{,\mu} = W^{(0)}_{,\mu} + \mathcal{O}(s)$. The Hamilton-Jacobi equation obtained from \eqref{eq:tdhamspec} then reads
\begin{align}
g^{\mu\nu} W^{(1\rm sw)}_{,\mu} W^{(1\rm sw)}_{,\nu}  - e^\kappa_{C;\nu}e_{D\kappa } s^{CD} W^{(0),\nu} + \mathcal{O}(s^2) =-1 \,, \label{eq:pertspjac}
\end{align}
where one should always remember that $s^{CD}$ is given by \eqref{eq:spparam} and $\cala\to W_{, \phi},\,\calb \to W_{,\psi}$.

When we further choose the signature of the background tetrad identical to that of $W^{(0)}_{,\mu}$ and the tetrad parameters $K_{\rm c},E_{\rm c},L_{\rm c}$ $\mathcal{O}(s)$-close  to the constants of motion of $W^{(0)}$, we obtain up to higher-order terms
\begin{align}
\begin{split}
&\Sigma g^{\mu\nu} W^{(1\rm sw)}_{,\mu} W^{(1\rm sw)}_{,\nu}  \\
&+ r^2 + 2 s^{12}\sqrt{K_{\rm c}} \frac{E_\mathrm{c} (r^2 + a^2) - a L_\mathrm{c}}{ r^2 + K_\mathrm{c}} \\
&+a^2 \csq \vth  +2 a s^{12}\sqrt{K_{\rm c}} \frac{L_\mathrm{c} - a E_\mathrm{c} \sin^2 \! \vartheta}{K_\mathrm{c} -  a^2 \cos^2 \! \vartheta}  = 0 \,. 
\end{split}
\end{align}
Now we notice that the only appearing component of spin is $s^{12} = \cala+\calb+s \to W_{,\phi} + W_{,\psi} + s$. The phases $\phi,\psi$ are thus cyclical coordinates and the initial values of $\cala,\calb$ integrals of motion. Furthermore, the Tulczyjew-Dixon condition boils down to $s^{C0}=0+ \mathcal{O}(s^2)$ or $W_{,\psi} = 0+ \mathcal{O}(s^2)$ (cf. eq. \eqref{eq:spparam}). Consequently, the value of the coordinate $\psi$ has no influence on either the spin tensor or the orbital motion at given order.

Finally, we can assume a separable action of the form $W^{(1\rm sw)} = -E_{\rm so} t + L_{\rm so} \vph +(s_\parallel - s)\phi + w_{r}(r) + w_\vth(\vth)$ with $E_{\rm so}, L_{\rm so}, s_{\parallel}$ some separation constants to obtain
\begin{align}
\begin{split}
(w'_\vartheta)^2 =& K_\mathrm{so} - \left( \frac{L_\mathrm{so}}{\sin \vartheta} - a E_\mathrm{so} \sin \vartheta \right)^2 - a^2 \cos^2 \! \vartheta \\  & -2a s_\parallel\sqrt{K_\mathrm{c}} \frac{L_\mathrm{c} - a E_\mathrm{c} \sin^2 \! \vartheta}{K_\mathrm{c} -  a^2 \cos^2  \! \vartheta}  \,,\end{split} \label{eq:wth}
\\
\begin{split}
\Delta (w'_r)^2 =& -K_\mathrm{so} + \frac{1}{\Delta} \left(E_\mathrm{so} (r^2 + a^2) - a  L_\mathrm{so} \right)^2 -  r^2 \\  & -2  s_\parallel \sqrt{K_\mathrm{c}} \frac{ E_\mathrm{c} (r^2 + a^2) - a L_\mathrm{c}}{K_\mathrm{c}+ r^2} \,, \end{split}\label{eq:wr}
\end{align}
where $K_{\rm so}$ is a separation constant analogous to the Carter constant. I discuss the meaning of the separation constants $K_{\rm so},E_{\rm so},L_{\rm so},s_{\parallel}$ in section \ref{subsec:constint}. At this point, I will only note that $K_{\rm c},E_{\rm c},L_{\rm c}$ only need to be chosen $\mathcal{O}(s)$ close to $K_{\rm so},E_{\rm so},L_{\rm so}$ for the Hamilton-Jacobi equation to be fulfilled up to $\mathcal{O}(s^2)$ terms in the swing region.

\subsection{Turning region corrections} \label{subsec:turn}
The swing solution of the perturbative action stays valid even when we shift $K_{\rm c},E_{\rm c},L_{\rm c}$ by an $\mathcal{O}(s)$ shift, and we can assume that we can always choose the congruence constants so that the motion corresponding to $W^{(1\rm sw)}$ avoids the turning points (and thus singularities) of the tetrad by an $\mathcal{O}(s)$ distance. 

Nevertheless, the connection terms diverge as $1/\sqrt{y - y_{\rm t}}$ near turning points $y_{\rm t}$, where either $y = r$ or $y= \cos \vth$. Additionally, $W^{(1\rm sw)}_{,y}$ becomes only $\mathcal{O}(\sqrt{s})$ close to the background congruence in those regions. Hence, at $y - y_{\rm t} \sim \mathcal{O}(s)$ several new terms contribute to the $\mathcal{O}(s)$ perturbation of the Hamilton-Jacobi equation. Let us write the equation in the respective $y$-turning region including all the $\mathcal{O}(s)$ terms as
\begin{align}
\begin{split}
&g^{\mu\nu} W^{(1)}_{,\mu} W^{(1)}_{,\nu}  - e^\kappa_{C;\nu}e_{D \kappa} s^{CD}e_0^\nu+1
\\
& \left[- e^\kappa_{C;y}e_{D \kappa} s^{CD}(W^{(1)}_{,y} -e_{0y}) + \frac{1}{4} (e_{C;y}^\kappa e_{D \kappa} s^{CD})^2\right] g^{yy} 
\\&=0 + \mathcal{O}(s^2) \,. 
\end{split}\label{eq:turnjac}
\end{align}
The first line in \eqref{eq:turnjac} corresponds to the swing-region terms, but the second line is new and makes $W^{(1 \rm sw)}$ an invalid solution for the action already at $\mathcal{O}(s)$.

I now use the following Ansatz in order derive turning-region corrections to the action. First, I replace all the instances of $s^{CD}$ with $\tilde{s}^{CD} \equiv s^{CD} (\cala \to s_\parallel-s; \calb \to 0)$ in the equations and assume that the solution using this Ansatz will be valid at least up to $\mathcal{O}(s)$. Next, I assume that there exist corrections of the form $\delta_{y}W^{(\rm t)} $ such that $W^{(1)} = W^{(1\rm sw)} + \delta_{r} W^{(\rm t)} + \delta_{\vth} W^{(\rm t)}$ and $\delta_{y} W^{(\rm t)} \sim s^2/\sqrt{y - y_{\rm t}}$. As a result, these corrections are of higher-order in the swing region and contribute to the Hamilton-Jacobi equation at $\mathcal{O}(s)$ only in their respective $y$-turning region. Then the corrections must fulfill
\begin{align}
\begin{split}
    & (\delta_{y } W^{(\rm t)}_{,y})^2 +2 w'_y \delta_{y} W^{(\rm t)}_{,y}
     \\
    &-  {e}_{C;y}^\kappa  {e}_{D \kappa} \tilde{s}^{CD}(w'_y  + \delta_{y} W^{(\rm t)}_{,y} - W^{(0)}_{,y})
    \\
    &
    + \frac{1}{4} ( {e}_{C;y}^\kappa  {e}_{D \kappa} \tilde{s}^{CD})^2 = 0\,.
\end{split}
\end{align}
Notice that thanks to the substitution $s^{CD} \to \tilde{s}^{CD}$ only derivatives with respect to $y$ appear in the equation.
The equation is solved by
\begin{align}
\begin{split}
    \delta_{y} W^{(\rm t)}_{,y} = 
    & -\left(w_y' - \frac{1}{2}  {e}_{C;y}^\kappa  {e}_{D \kappa} \tilde{s}^{CD}\right) 
    \\
    & \pm\sqrt{w_y'^2 - e_{0 y} {e}_{C;y}^\kappa  {e}_{D \kappa} \tilde{s}^{CD}}\,, \label{eq:deltW}
\end{split}
\end{align}
where I have used the fact that up to higher orders $W^{(0)}_{,y} = e_{0 y}$ to simplify notation. We now notice that the expressions for $\delta_{y} W^{(\rm t)}_{,y}$ are not separable, so we have to write
\begin{align}
    &\delta_{r} W^{(\rm t)} = \int \delta_{r} W^{(\rm t)}_{,r} \di r + C_{r}(\phi,\vartheta)\,, \\
    &\delta_{\vth} W^{(\rm t)} = \int \delta_{r} W^{(\rm t)}_{,\vartheta} \di \vartheta + C_{\vartheta}(\phi,r)\,.
\end{align}
Particular choices of $C_y$ might improve the properties of the approximation, but I will use here $C_y = 0$. 

Finally, by substituting the action including the turning-region corrections back in the full Hamilton-Jacobi equation we see that the error terms are $\mathcal{O}(s^2)$ terms in the swing regions, and $\mathcal{O}(s^{3/2})$ in the turning regions. Finding necessary turning-point corrections even for these $\mathcal{O}(s^{3/2})$ terms is necessary before terms quadratic in spin can be included. However, I will leave this task for further work. 

In summary, the action valid both in the turning and swing regions up to $\mathcal{O}(s)$ terms reads
\begin{align}
\begin{split}
    W^{(1)}(t,\varphi,r,\vartheta,\phi) =& (s_\parallel - s) \phi -E_{\rm so} t + L_{\rm so} \varphi \\ 
    + \sum_{y=r,\vartheta} \int \Big(& \pm \sqrt{w_y'^2 -   e_{0y}{e}_{C;y}^\kappa  {e}_{D \kappa} \tilde{s}^{CD}} \\ 
    & +  \frac{1}{2}  {e}_{C;y}^\kappa  {e}_{D \kappa} \tilde{s}^{CD}  \Big)\di y \,, \label{eq:wfull}
\end{split}
\end{align}
where $s_\parallel,K_{\rm so},E_{\rm so},L_{\rm so}$ are to be understood as parameters of the family of solutions, $w'_r, w'_\vth$ are given in equations \eqref{eq:wr} and \eqref{eq:wth}, and $\tilde{s}^{CD} = -\tilde{s}^{DC}$ is explicitly given as $\tilde{s}^{0D} = 0,\,\tilde{s}^{12} = s_\parallel,\tilde{s}^{23} = \sqrt{s^2 - s_\parallel^2}\sin \phi,\,,\tilde{s}^{31} = \sqrt{s^2 - s_\parallel^2}\cos \phi$.

\subsection{Interpretation of separation constants} \label{subsec:constint}

Let us define the following orbital functions corresponding to orbital Carter constant, specific energy, and specific angular momentum
\begin{align}
    &K_{\rm o} \equiv u_\vth^2 + \left(\frac{u_\vph}{\sin \vth} + a u_t \sin \vth \right)^2 +a^2 \csq \vth\,,\\
    &E_{\rm o} \equiv -u_t \,,\\
    &L_{\rm o} \equiv u_\vph\,.
\end{align}
For geodesics, these orbital functions are constant whereas for the spin-perturbed orbit they are not. To show that, I use the fact that
\begin{align}
    u_\mu = W_{,\mu} - \frac{1}{2} e_{C;\mu}^\kappa e_{D \kappa} s^{CD}\,.
\end{align}
Now one can relate the spin-orbital constants of motion $K_{\rm so},E_{\rm so}, L_{\rm so}$ to the orbital functions defined above as
\begin{align}
&E_\mathrm{so} = E_\mathrm{o} + \frac{1}{2} \Gamma_{CD t}  \tilde{s}^{CD} +\mathcal{O}(s^2;s^{3/2})\,, \label{eq:EoEso}\\
&L_\mathrm{so} = L_\mathrm{o} - \frac{1}{2} \Gamma_{CD \varphi} \tilde{s}^{CD} +\mathcal{O}(s^2;s^{3/2})\,, \label{eq:LoLso}\\
\begin{split}
&K_\mathrm{so} 
= K_\mathrm{o} -   e_{0 \vartheta} {e}_{C \kappa}  {e}^\kappa_{D;\vartheta} \tilde{s}^{CD} 
\\
&+ \left(\frac{L_\mathrm{o}}{\sin \vartheta} - a E_\mathrm{o} \sin \vartheta\right)\left(\frac{\Gamma_{CD \varphi}}{\sin \vartheta} + a \sin \vartheta \Gamma_{CD t} \right) \tilde{s}^{CD} \\& +2  a\sqrt{K_\mathrm{o}} \frac{L_\mathrm{o} - a E_\mathrm{o} \sin^2 \! \vartheta}{K_\mathrm{o} - a^2 \cos^2  \! \vartheta} s_\parallel+\mathcal{O}(s^2;s^{3/2}) \,, \label{eq:KoKso}
\end{split}
\end{align}
where $\mathcal{O}(s^2;s^{3/2})$ denotes error terms of order $\mathcal{O}(s^2)$ in the swing regions and $\mathcal{O}(s^{3/2})$ in the turning regions. Furthermore, $\Gamma_{CD\kappa} \equiv \Gamma_{\mu\nu\kappa}  {e}^\mu_D  {e}^\nu_D$ are the Christoffel symbols projected into the tetrad (I used the fact that the tetrad is independent of $t,\varphi$). 

Now we see that the left-hand sides of equations \eqref{eq:EoEso}, \eqref{eq:LoLso} and \eqref{eq:KoKso} are constant and the right hand sides contain $K_{\rm o},E_{\rm o}, L_{\rm o}$ and fluctuating terms, which makes the orbital energy, angular momentum, and ``orbital Carter constant'' time-variable. Furthermore, it is easy to see that the spin-orbital constants of motion are generally not equal even to average values of the orbital functions and that there is a persistent $\mathcal{O}(s)$ shift between the two.

Constants of motion of spinning particles under the Tulczyjew-Dixon condition were studied by \citet{ruediger1,ruediger2}. He found exact integrals of motion in space-times with an explicit symmetry and a corresponding Killing vector $\xi^\mu$ given as
\begin{align}
    C_{\xi} = u_\mu\xi^\mu - \frac{1}{2} \xi_{\mu;\nu} s^{\mu\nu}\,.
\end{align}
By comparing these integrals of motion corresponding to the $t$ and $\varphi$ symmetry of Kerr space-time, we find that they are exactly equal to the constants $-E_{\rm so}, L_{\rm so}$ respectively. 

Additionally, R{\"u}diger found approximately conserved quantities associated with the existence of a Killing-Yano tensor $Y_{\mu\nu}$, which can be written as
\begin{align}
    & C_\mathrm{Y} = \frac{1}{2}Y_{\mu\nu} u^\nu \varepsilon^{\mu\kappa\lambda\gamma}u_\kappa s_{\lambda \gamma}\,, \\
    & K_\mathrm{R} = Y_{\mu\chi} Y_{\nu}^{\;\;\chi} u^\mu u^\nu - 2 u^\mu s^{\rho \sigma} \left(Y_{\mu\rho;\kappa} Y^{\kappa}_{\;\;\sigma} + Y_{\rho \sigma;\kappa} Y^{\kappa}_{\;\;\mu}\right)\,, \label{eq:KR}
\end{align}
where $\dot{K}_{\rm R} = \mathcal{O}(s^2), \dot{C}_{\rm Y} = \mathcal{O}(s^2)$. The interpretation of $C_{\rm y}$ is that of the projection of the specific spin vector $s^\mu = \epsilon^{\mu\nu\kappa\lambda} s_{\nu\kappa}u_\lambda/2$ into the specific angular momentum vector $l_\mu = Y_{\mu\nu}u^\nu$, $C_Y = s^\mu l_\mu$. The constant $K_{\rm R}$ can be loosely interpreted as some sort of ``spin-orbital angular momentum squared''. However, notice that $K_{\rm R}$ is {\em not} the square of the vector $l^\mu + s^\mu$.

Now we can compare with the separation constants such as $s_\parallel, K_{\rm so}$. Let us compute
\begin{align}
\begin{split}
    s_\parallel &= 
    \tilde{s}^{12} = \frac{1}{2}\varepsilon^{30CD}\tilde{s}_{CD} = \frac{1}{2} e_{3\mu} e_{0\kappa}\varepsilon^{\mu \kappa \lambda \gamma}\tilde{s}_{\lambda \gamma}
    \\&= \frac{C_{Y}}{\sqrt{K_{\rm c}}} + \mathcal{O}(s^2;s^{3/2})\,.
\end{split}
\end{align}
In other words, $C_{\rm Y} = \sqrt{K_{\rm c}}s_\parallel+ \mathcal{O}(s^2,s^{3/2})$. Notice that it is exactly this factor that appears in $w'_y$ and we can thus say that the correction to the action in the swing region is proportional to $l_\mu s^\mu$. Furthermore, the cases $s_\parallel=\pm s$ correspond to $\tilde{s}^{23} =\tilde{s}^{31}=0$ and the spin completely aligned or counter-aligned with the orbital angular momentum respectively.  

To compare $K_{\rm R}$ with $K_{\rm so}$, one simply needs to substitute the spin-perturbed four-velocity into \eqref{eq:KR}, and a somewhat involved computation yields
\begin{align}
    K_{\rm R} = K_{\rm so}  + \mathcal{O}(s^2;s^{3/2})\,.
\end{align}
In summary, the separation constants of motion are in a straightforward relation with those of \citet{ruediger1,ruediger2}, which is an important consistency check for the solution \eqref{eq:wfull}.  


\section{Orbital motion} \label{sec:orb}
In many problems in classical mechanics, orbital motion turns out to be separable and solvable by a finite set of quadratures (closed-form integrals) once the Hamilton-Jacobi equation has been separated. However, we will see that the perturbative construction of the action from the last section does not allow for such a separation of orbital equations of motion.

Nevertheless, in section \ref{subsec:turnpoints} I will show that it is still possible to analytically determine the corrections to the turning points of the motion, and, in return, this is used in section \ref{subsec:freq} to determine the corrections to the fundamental frequencies of motion by a finite set of quadratures.

\subsection{Equations of motion}
The equations of motion for the spin-perturbed trajectory in Mino time read
\begin{subequations}
\label{eq:vel}
\begin{align}
 \frac{\di r}{\di \lambda} = \pm&\Delta\sqrt{{w'_r}^2 - e_{0r}  {e}^\kappa_{C;r} {e}_{\kappa B}  \tilde{s}^{CD} } \,, \\
 \frac{\di \vartheta}{\di \lambda} = \pm&  \sqrt{{w'_\vartheta}^2 -  e_{0 \vartheta}  {e}^\kappa_{C;\vartheta} {e}_{\kappa B}  \tilde{s}^{CD} }\,, \\
\begin{split}
\frac{\di \phi}{\di \lambda} =  
-&\sqrt{K_\mathrm{c}} \Big( \frac{E _\mathrm{c} (r^2 + a^2) - a L _\mathrm{c}}{K_\mathrm{c}+r^2}
\\
&\quad\quad\quad+ a \frac{L _\mathrm{c} - a E _\mathrm{c} \sin^2 \! \vartheta}{K _\mathrm{c} - a^2 \cos^2 \! \vartheta} \Big)\,, \label{eq:phipr}
\end{split}
\end{align}
\end{subequations}
where I have discarded $\mathcal{O}(s^2;s^{3/2})$ terms in the $r,\vth$ equations, and $\mathcal{O}(s,\sqrt{s})$ terms in the $\phi$ equation. Such a term-discarding scheme is consistent with the accuracy to which the Hamilton-Jacobi equation was solved as well as with the goal to acquire $r(\lambda),\vth(\lambda)$ orbital shape at $\mathcal{O}(s)$ precision. We immediately see that the equations of motion are not separable, since the connection terms are mixed and since $\tilde{s}^{CD}$ involves trigonometric functions of $\phi$.

One interesting feature of the equations of motion is the change in the symmetries as compared to geodesic motion. For instance, $|\di \vth/\di \lambda|$ is not symmetric with respect to reflections about the equatorial plane $\vth \to \uppi - \vth$; it is only symmetric with respect to the combined transformation consisting of a reflection $\vth \to \uppi - \vth$ coupled with either $\tilde{s}^{CD}\to -\tilde{s}^{CD}$ or $\di \vth/\di \lambda \to -\di \vth/\di \lambda$. In other words, when the particle is at a given distance $|\vth-\uppi/2|$ from the equatorial plane, it will move at a slightly different $\di \vth/\di \lambda$ when it is moving \textit{towards} the equatorial plane than when it is moving \textit{away} from the equatorial plane.

\subsection{Turning points} \label{subsec:turnpoints}

\begin{figure*}
    \centering
    \includegraphics[width=0.4\textwidth]{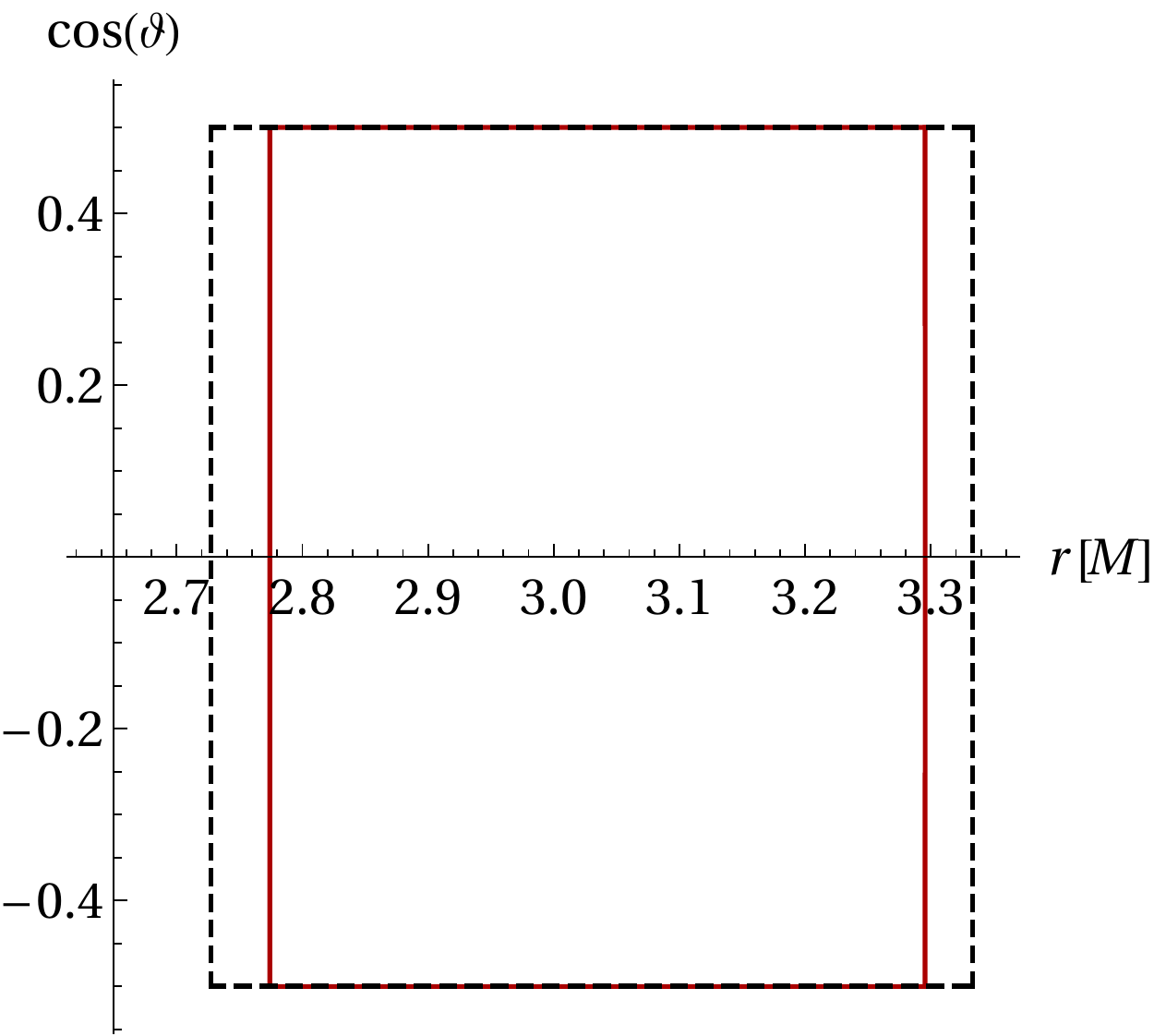} \includegraphics[width=0.4\textwidth]{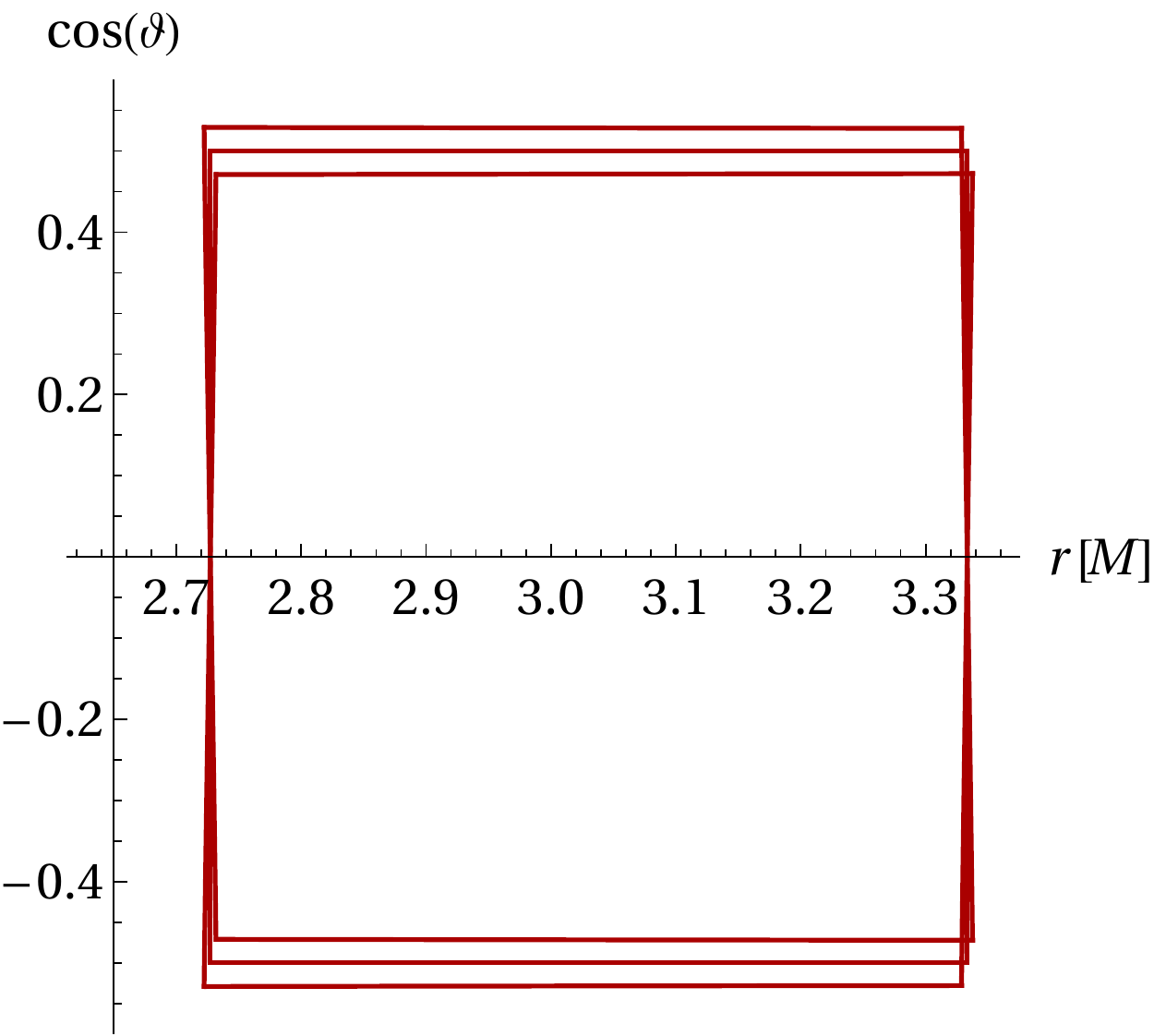}
    \caption{The turning points of particles near a Kerr black hole with $a=0.9M$ and with various choices of particle spin. The fiducial geodesic is always with constants $K = 2.9 M^2, E = 0.87 , L = 2.0 M$ (or semi-latus rectum $p=3M$, eccentricity $e = 0.1$, and inclination $\sqrt{z_{-\rm g}} = 0.5$). On the left we plot the turning points of the fiducial geodesic (dotted black) and the turning points of a spin-perturbed orbit with completely aligned spin, $s_\parallel = s = 10^{-3} M$ (full dark red). On the right we show the turning points of an orbit with a completely oscillating spin $s_\parallel=0, s= 5\cdot 10^{-2}   M$ for the values of spin angle $\phi=0,\uppi/2,3\uppi/2$.  }
    \label{fig:turnboxes}
\end{figure*}
The condition for the turning points can be written as
\begin{align}
    (w'_y)^2 -  e_{0y}  {e}_{C;y}^\kappa  {e}_{D \kappa} s^{CD} = 0\,,
\end{align}
where again $y=r,\vth$. Now it might be tempting to discard the connection term since $e_{0y} \sim \mathcal{O}(\sqrt{s})$ near the turning point, but the $e_{C;y}^\kappa$ will counter it by becoming $\mathcal{O}(1/\sqrt{s})$ in the turning region. Thus, the whole connection term stays $\mathcal{O}(s)$ and needs to be included in the computation.

I express the turning points in terms of shifts with respect to turning points of fiducial geodesics. It is important to note that these fiducial geodesics will {\em not} be the same geodesics as the ones generating the tetrad congruence. Specifically, I choose the fiducial geodesic to have constants of motion $K=K_\mathrm{so} - 2 a s_\parallel \mathrm{sgn}(L_\mathrm{so} - a E_\mathrm{so})$, $L=L_\mathrm{so}$, and $E=E_\mathrm{so}$. Then I assume that the turning point is $\mathcal{O}(s)$-shifted away from the turning point of the geodesic, $y_{\rm t} = y_{\rm gt} + \delta y_{\rm t}, \, \delta y_{\rm t} \sim \mathcal{O}(s)$. When the dust settles, I obtain
\begin{subequations} \label{eq:turnshifts}
\begin{align}
     \delta r_\mathrm{t} =& \frac{2 s_\parallel \mathcal{G} + \Delta (K + r^2 )  \mathcal{X}^{(r)}_{\kappa C} e_{D}^\kappa \tilde{s}^{CD}}{\mathcal{I}_{(r)} (K +  r^2)}\Big|_{r=r_\mathrm{gt}} \!\!, \label{eq:deltart}
    \\
     \delta \vartheta_\mathrm{t} =& \frac{2as_\parallel \mathcal{H} + (K  -   a^2 \cos^2\! \vartheta) \mathcal{X}^{(\vartheta)}_{\kappa C} e_{D}^\kappa \tilde{s}^{CD}}{\mathcal{I}_{(\vartheta)}(K  -   a^2 \cos^2\! \vartheta)}\Big|_{\vartheta=\vartheta_\mathrm{gt}} \!\!, \label{eq:deltatht}
    \\
    \begin{split}
    \mathcal{G} \equiv& \sqrt{K}(E(r^2 + a^2) -a L ) \\
    &+ a (K +   r^2) \mathrm{sgn}(L-aE)\,,
    \end{split}\\
    \mathcal{I}_{(r)} \equiv&  \frac{\di}{\di r}\left(\Delta^{-1} \left[E (r^2 + a^2) - a L \right]^2 -   r^2\right) \,, \\
    \begin{split}
     \mathcal{H} \equiv& \sqrt{K } (L  - a E  \sin^2\!\vartheta) 
    \\ & - (K  -   a^2 \cos^2\! \vartheta) \mathrm{sgn}(L-aE) \,,
    \end{split}\\
     \mathcal{I}_{(\vartheta)} \equiv & -\frac{\di}{\di \vartheta} \left[ \left(L - a E \sin^2\! \vartheta \right)^2 \sin^{-2} \! \vartheta +   a^2 \cos^2 \! \vartheta \right] \!,
\end{align}
\end{subequations}
where $\mathcal{X}^{(y)}_{\kappa D} \equiv \lim_{y\to y_{\rm gt}}  e_{0y} e_{D \kappa;y}$. The coefficients $\mathcal{X}^{(y)}_{\kappa D}$ are then easy to compute as
\begin{subequations} \label{eq:turnX}
\begin{align}
    & \mathcal{X}^{(r)}_{\kappa 0} \di x^\kappa = \frac{ \mathcal{I}_{(r)}}{2   \Delta}  \di r \,,
    \\
    & \mathcal{X}^{(r)}_{\kappa 1} \di x^\kappa =   \frac{\alpha \mathcal{I}_{(r)}}{2  \Delta}  \di r\,,
    \\
    & \mathcal{X}^{(r)}_{\kappa 2} \di x^\kappa =  - \frac{\alpha r  \mathcal{I}_{(r)}}{2   \Sigma \sqrt{K}} ( \di t - a \sin^2\! \vartheta  \di \varphi)\,,
    \\
     & \mathcal{X}^{(r)}_{\kappa 3} \di x^\kappa = - \frac{a \cos \vartheta  \mathcal{I}_{(r)}}{2  \Sigma \sqrt{K}} ( \di t - a \sin^2\! \vartheta  \di \varphi)\,,
     \\
     & \mathcal{X}^{(\vartheta)}_{\kappa 0} \di x^\kappa = \frac{\mathcal{I}_{(\vartheta)}}{2  }  \di \vartheta \,,
     \\
     & \mathcal{X}^{(\vartheta)}_{\kappa 1} \di x^\kappa =\frac{1}{\alpha} \frac{\mathcal{I}_{(\vartheta)}}{2 }  \di \vartheta\,,\\
       & \mathcal{X}^{(\vartheta)}_{\kappa 2} \di x^\kappa = -\frac{1}{\alpha} \frac{a \cos \vartheta \sin \vartheta \mathcal{I}_{(\vartheta)}}{2  \Sigma \sqrt{K}} (a  \di t - (r^2 + a^2)  \di \varphi) \,,
       \\
     & \mathcal{X}^{(\vartheta)}_{\kappa 3} \di x^\kappa = \frac{r \sin\vartheta \mathcal{I}_{(\vartheta)}}{2  \Sigma \sqrt{K}} (a  \di t - (r^2 + a^2)  \di \varphi) \,,
     \\
     & \alpha \equiv \sqrt{\frac{K - a^2 \cos^2 \! \vartheta}{K + r^2}}\,,
\end{align}
\end{subequations}
where the expressions for $\mathcal{X}^{(y)}_{\kappa D}$ are evaluated at the respective $y=y_{\rm gt}$. Note that I have discarded $\mathcal{O}(s^{3/2})$ terms that come from the fact that the tetrad-congruence constants $K_{\rm c},E_{\rm c},L_{\rm c}$ are generally $O(s)$-shifted with respect to the fiducial-geodesic constants $K,E,L$. In the Appendix I will not discard such terms for technical reasons and the term $\mathcal{X}^{(y)}_{\kappa C} e_{D}^\kappa$ is then simply replaced by $e_{0y} e_{C \kappa;y}e_{D}^\kappa|_{y=y_{\rm gt}}$.

To perform a consistency check of the results, I have taken the effective potentials of spinning particles with aligned spin in the equatorial plane as given by \citet{tod1976} or \citet{hackmann2014}, and found its turning points to linear order in spin to obtain the exact same results as in \eqref{eq:turnshifts}. 

One should notice that the formulae \eqref{eq:turnshifts} are finite for motion in the equatorial plane ($K = (L-a E)^2$) only thanks to the choice $K=K_\mathrm{so} - 2 a s_\parallel \mathrm{sgn}(L_\mathrm{so} - a E_\mathrm{so})$. For other choices of the fiducial mapping, the $\vartheta$ turning points of spin-perturbed near-equatorial motion will be $\mathcal{O}(\sqrt{s})$ far from the turning points of the fiducial geodesic. 

However, the choice of the fiducial geodesics that I give here does not yet avoid similar singularities in $\delta r_{\rm t}$ as the motion becomes near-circular. To regularize this case, one must choose $E(E_{\rm so}, L_{\rm so}),L(E_{\rm so}, L_{\rm so})$ so that the fiducial geodesic circularizes at an $\mathcal{O}(s)$-close radius and for the same values of $E_{\rm so}, L_{\rm so}$ as the spin-perturbed orbit itself. Such a construction is somewhat involved and I leave it for future work. 

Let us now shortly discuss the qualitative features of the spin-perturbed turning points. The first thing to notice is the fact that the turning points are not separable any more, the connection terms are functions of both $r,\vth$, and $\tilde{s}^{CD}$ is generally a function of $\phi$ whenever $s_\parallel \neq \pm s$. The shape of the ``turning box'' in the $r,\vth$ plane is illustrated in Fig. \ref{fig:turnboxes}. One last thing to notice is the fact that in the aligned/counter-aligned case $s_\parallel = \pm s$ the spin tensor does not depend on $\phi$ and the turning box is symmetric about the equatorial plane. In general, however, the shifts of the turning points are invariant with respect to transformations $\vth \to \uppi - \vth$ only in combinations with either $\phi \to -\phi$ or $\phi \to \phi + \uppi$.

\subsection{Spin-orbital actions and resonances} \label{subsec:actions}
The relations $\mathcal{U}_y = \partial W^{(1)}(x,y,\phi;...)/\partial y$, $\mathcal{A} = \partial W^{(1)}(x,y,\phi;...)/\partial \phi$ define a three-torus $\mathbb{T}^3$ to which the motion is identified in the dynamical part of the phase space $(r,\vartheta,\phi,\mathcal{U}_r,\mathcal{U}_\vartheta,\mathcal{A})$, at least in the case of bound motion that is of interest here. Whenever such a torus identified in a dynamical system with canonical coordinates $p_i,q^i$, it is possible to define the action $I_\gamma$ over a loop $\gamma$ on the torus as
\begin{align}
    I_{\gamma} \equiv \frac{1}{2 \uppi} \oint_\gamma \!\! p_i \mathrm{d} q^i\,,
\end{align} 
where the generalized Stokes theorem implies that this integral is only dependent on the homotopy equivalence class of the loop (the class of loops that can be deformed into each other without discontinuing the loop).
When, furthermore, the actions are defined over $n$ homotopically inequivalent loops over an $n$-torus $\mathbb{T}^n$, they can always be completed into an action-angle system of coordinates \cite{arnold2007}.

I now define the set of actions 
\begin{subequations} \label{eq:actions}
\begin{align}
    & I_y = \frac{1}{\uppi}\int_{y_{\rm t1}}^{y_{\rm t2}} \!\! \frac{\partial W^{(1)}}{\partial y} \di y \Big|_{x,\phi = \rm const.}\,,\\
    & I_\phi =  \frac{1}{2 \uppi}\int_0^{2 \uppi} \!\! \frac{\partial W^{(1)}}{\partial \phi} \di \phi \Big|_{x,y = \rm const.}\,,
\end{align}
\end{subequations}
where the integration bounds $y_{\rm t 1,2}$ are the turning points computed in the last subsection. The value of the actions $I_y,I_\phi$ is dependent only on the constants of motion $K_{\rm so}, E_{\rm so}, L_{\rm so}, s_{\parallel}$ and not on the constant values of the coordinates we are not integrating over (different values of these coordinates represent homotopy-equivalent integration loops). It is now simple to show that the actions fulfill
\begin{align}
    \frac{\mathrm{d} I_{y,\phi}}{\mathrm{d} \lambda} = 0 + \mathcal{O}(s^{2}) \,. \label{eq:Icons}
\end{align}
This also proves that, to linear order in spin, the spin perturbation will not cause any topological changes in the action foliation of the phase space around resonant orbits (in contrast to generic perturbed integrable systems). The recent numerical study of Ref. \cite{zelenka2019} has examined several resonances caused by the spin and only found those caused by second order in spin.

\subsection{Fundamental frequencies} \label{subsec:freq}

\begin{figure}
    \centering
    \includegraphics[width=0.48\textwidth]{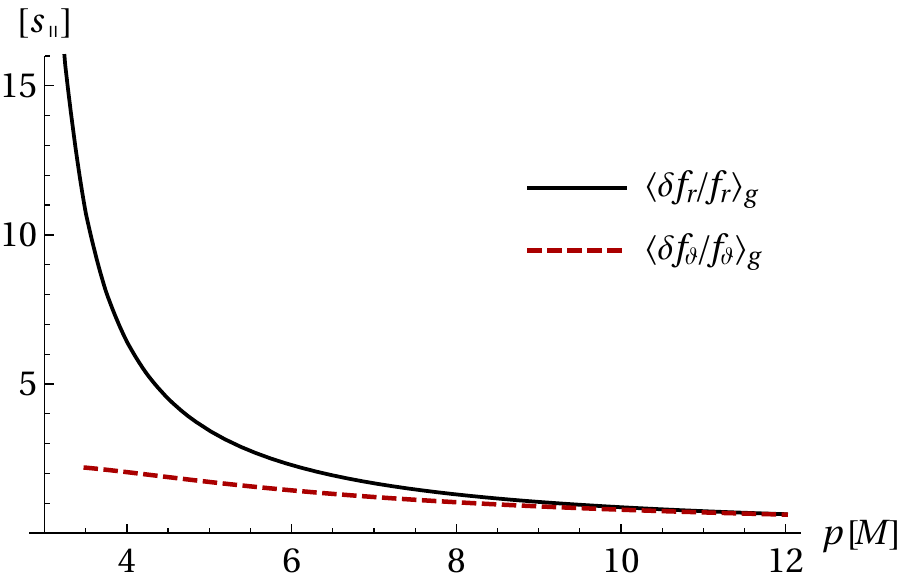}
    \caption{The relative corrections to fundamental frequencies given in units of the aligned component of spin $s_\parallel$ as a function of semi-latus rectum $p$ for fiducial geodesics with $e=0.1$, $z_{- \rm g} = 0.1$, and $a=0.9 M$. The corrections to the radial frequency becomes large at small $p$ because the motion is becoming unstable.}
    \label{fig:freqs}
\end{figure}

One of the main issues with the computation of fundamental frequencies and various averages over the spin-perturbed motion is the inseparability of the turning points. Consequently, it is not even clear which integration bounds should be chosen in the computations. I resolve this issue by transforming to a set of angle-type coordinates $(\chi_r,\chi_\vartheta) \in (0,2\uppi]^2$ such that
\begin{subequations} \label{eq:angparam}
\begin{align}
    & r(\chi_r,\chi_\vartheta,\phi) = r_0+ \delta r \,, \label{eq:rparam}\\
    & \cos \left[ \vartheta(\chi_r,\chi_\vartheta,\phi) \right] = \cos \vartheta_0 - \delta \vartheta \sqrt{1-z_{-\rm g}} \,,\label{eq:thparam}\\
    & r_0 \equiv \frac{r_{1\rm g}  + r_{2\rm g}}{2} + \frac{r_{1\rm g}  - r_{2\rm g}}{2} \sin\chi_r \,,\\
    & \delta r \equiv \frac{\delta r_{1}  + \delta r_{2}}{2} + \frac{\delta r_{1}  - \delta r_{2}}{2} \sin\chi_r \,,\\
    & \cos\vartheta_0  \equiv \sqrt{z_{-\rm g}} \sin\chi_\vartheta \,,\\
    & \delta \vartheta \equiv \frac{\delta \vartheta_{1}  + \delta \vartheta_{2}}{2} + \frac{\delta \vartheta_{1}  - \delta \vartheta_{2}}{2} \sin\chi_\vartheta\,,\\
    & \delta y_i(\chi_x,\phi) \equiv \delta y_{\rm t} (y_{\rm g i},x_0(\chi_x),\phi)\,,
\end{align}
\end{subequations}
where $i=1,2$ and $\{x,y\} = \{r,\vartheta\},\{\vartheta,r\}$. All the details of the computations using this transformation are discussed in the Appendix. When the dust settles, the equations of motion reduce to
\begin{align}
    & \frac{\di \chi_y}{\di \lambda} = f_{y}(\chi_y) + \delta f_{y}(\chi_r,\chi_\vartheta,\phi)\,,\\
    & \frac{\di \phi}{\di \lambda} = h(\chi_r,\chi_\vth)\,. \label{eq:phip}
\end{align}
The functions $f_y$ are $\mathcal{O}(1)$, non-zero, and regular for all $\chi_y$. On the other hand, $\delta f_y$ are $\mathcal{O}(s)$ and mostly regular with ignorable singular terms at turning points. 

Any state of the spin-perturbed trajectory can now be specified by some point in the compact phase space $(\chi_r,\chi_\vartheta,\phi) \in (0,2\uppi]^3$ and one can apply usual perturbation and averaging procedures accordingly. When the dust settles, the fundamental Mino angular frequencies of the system of equations turn out to be
\begin{subequations} \label{eq:funfreqs}
\begin{align}
    &\Upsilon_{r} = \Upsilon_{r\rm g}\left(1 +   \left\langle \frac{\delta f_r}{f_r} \right\rangle_{\!\!\rm g} \right) \,,\\
    &\Upsilon_{\vartheta} = \Upsilon_{\vartheta\rm g}\left(1 +  \left\langle \frac{\delta f_\vartheta}{f_{\vartheta}} \right\rangle_{\!\!\rm g} \right) \,,\\
    & \left\langle j(\chi_r,\chi_\vartheta,\phi)\right\rangle_{\rm g} \equiv \frac{\Upsilon_{r \rm g} \Upsilon_{\vartheta \rm g}}{(2 \uppi)^{3}}  \int_{(0,2\uppi]^3} \frac{j}{f_r f_\vartheta} \di \chi_r \di \chi_\vartheta \di \phi\,,
\end{align}
\end{subequations}

where $\langle j\rangle_{\rm g}$ means averaging the function $j$ over the fiducial geodesic. Note that since all the relevant expressions are linear in spin and since all the other components of the spin tensor are fully oscillating or zero, only the $\tilde{s}^{12} = -\tilde{s}^{21} = s_\parallel$ aligned component of spin survives in any geodesic average. In other words, only the value of $s_\parallel$ (and not of $s$) is important in the long-term effects of the spin perturbation. 

I have computed the relative frequency shifts $\langle \delta f_y/f \rangle_{\rm g}$ for a sample of geodesics and plotted them in Fig. \ref{fig:freqs}. There is nothing particularly unexpected about the qualitative behavior of the shifts; they are a factor of few times the spin and the radial corrections diverge as the motion becomes radially unstable near the black hole. 

One can also use this formalism to compute the average azimuthal angular frequency $\bar{\Upsilon}_\varphi$ and average rate of coordinate time with respect to Mino time $\Xi$ to obtain
\begin{align}
\begin{split}
    &\Xi \equiv \left\langle \frac{\di t}{\di \lambda}\right\rangle_{\rm \!\! sp} =
    \left\langle\frac{(r^2 + a^2) \mathcal{J}}{\Delta} \right\rangle_{\rm \!\! sp} 
   \\& \quad\quad\quad\quad\quad \quad\; - a E_{\rm so} \left\langle \sin^2 \vartheta\right\rangle_{\rm  sp} +  a L_{\rm so}
   \\& \quad\quad\quad\quad\quad \quad\;- \langle \Gamma_{CD}^{\quad\;t}\,\tilde{s}^{CD} \rangle_{\rm g}
   \,,
    \end{split}\\
    \begin{split}
    &\bar{\Upsilon}_\varphi \equiv \left\langle \frac{\di \varphi}{\di \lambda}\right\rangle_{\rm \!\! sp} =  \left\langle\frac{a\mathcal{J}}{\Delta} \right\rangle_{\rm \!\! sp} + \left\langle\frac{L_{\rm so}}{\sin^2 \vartheta}\right\rangle_{\rm \!\! sp} -  a E_{\rm so} 
    \\& \quad\quad\quad\quad\quad \quad\quad- \langle \Gamma_{CD}^{\quad\;\varphi}\,\tilde{s}^{CD} \rangle_{\rm g} \,,
\end{split} 
\\  &\mathcal{J}\equiv E_\mathrm{so}(r^2 + a^2) - a L_\mathrm{so}\,,
\end{align}
where $\langle \rangle_{\rm sp}$ means averaging over the spin-perturbed orbit. In general, one needs to know the shape of the spin-perturbed orbit for such averaging. However, all the ``sp'' averages we need to compute above are of separable functions, for which we can use (see section \ref{app:spavg} in the Appendix)
\begin{align}
\begin{split}
    \langle n(y) \rangle_{\rm sp}  =& \left(1+\left\langle \frac{\delta f_y}{f_y} \right\rangle_{\rm \!\!g} \right)\langle n(y_0) \rangle_{\rm g} + \langle n'(y_0) \delta y \rangle_{\rm g}
    \\& - \left\langle n(y_0) \frac{\delta f_y}{f_y} \right\rangle_{\!\! \rm g}\,.
\end{split} \label{eq:spavg}
\end{align}
Finally, the average coordinate-time angular frequencies are $\bar{\Omega}_y = \Upsilon_y/\Xi, \bar{\Omega}_\varphi = \bar{\Upsilon}_\varphi/\Xi$\,. It is also possible to compute the average angular frequency of the spin phase $\phi$ by the same methods as above, but this frequency will not be observable in any signal at leading order.

\section{Discussion and outlooks} \label{sec:disc}

{\em Hidden symmetry and multipole particles.} The separation of variables of the Hamilton-Jacobi equation in the swing region is a consequence of the hidden symmetry of the Kerr space-time and so is the conservation of $K_{\rm so}$. However, this result may be puzzling for the following reason. On one hand, the pole-dipole MPD equations describe the motion of systems that are keeping in balance by internal exchange of momentum (e.g. neutron stars), and on the other hand, it was shown that the conservation of the sum of Carter constants is violated for any system with components that exchange momentum \citep{grant2015,witzany2017}.

This apparent discrepancy is easily explained; the pole-dipole system of equations is universal in the sense that it represents and evolves any body of the given value of the mass multipoles in the same way. In other words, an initially compact rotating cloud of non-interacting free test particles can be described in a multipolar expansion, and the equations of motion truncated at the pole-dipole level will be the same as the ones we use for an astrophysical compact body -- apart from the fact that the cloud will spread and the higher-order multipoles quickly become non-negligible. Since the free-streaming cloud {\em does} conserve the sum of its particles' Carter constants, an approximate ``total Carter constant'' such as $K_{\rm R}, K_{\rm so}$ {\em must} exist for the pole-dipole system of equations. 

Nevertheless, the pole-dipole order is the only order where we can replace an astrophysical body with a cloud of dust and obtain the same equations of motion; the pole-dipole-quadrupole equations are not universal anymore since the quadrupole dynamics include the composition-specific response of the body \citep{steinhoff2012,vines2016}. While the cloud of free-streaming particles should still conserve a total Carter constant in the multipolar formalism, there is no reason to believe this will be the case for the astrophysical body described by different evolution equations. On the contrary, since the quadrupole dynamics of the astrophysical body are governed by momentum-exchanging processes in its interior, one should expect no conservation of the sum of Carter constants. 

Thus, I believe that the description of general classical bodies to pole-dipole order is {\em precisely} the point to which hidden symmetry is relevant. In other words, I believe it will not be possible to generalize the construction given in this paper to higher order multipoles and powers of spin, at least for the motion of objects with sufficiently general tidal response. 

On the other hand, one can speculate that special cases of interest might still exhibit a conserved Carter-like constant even at higher (or even all) multipole orders. Isolated black holes possess a specific tower of Geroch-Hansen mass multipoles that are all generated by the black hole spin \citep{hansen1974}. However, a major obstruction in translating this to the MPD equations is the fact that there is no established dictionary between the Geroch-Hansen and MPD multipoles on a generally curved background (on a near-flat background, the case seems reasonably clear \citep{vines2018}). Specifically, if we postulate the multipoles under one supplementary spin condition and transform to a different one, the multipoles will transform as well (see, e.g, Ref. \citep{vines2016}). So the question is: In which frame are the MPD multipoles of a dynamical black hole the Geroch-Hansen multipoles? Where is the centroid of a spinning black hole moving in a general background space-time? 

{\em Resonances and chaos.} The perturbative solution of the Hamilton-Jacobi equations as presented in this paper does not rely on the orbits having non-integer ratios between frequencies and, thus, there is no reason to believe the solution is problematic around resonant orbits, as is also indicated in section \ref{subsec:actions}. 

 On the other hand, the action $W^{(1)}$ can be fully valid while still producing resonant effects on the level of the orbital shapes generated by the equations of motion \eqref{eq:vel}. For instance, the perturbative computation of the fundamental frequencies in \eqref{eq:funfreqs} will be ill-defined when the motion is resonant and $\delta f_y/f_y$ has a corresponding non-zero harmonic. \citet{ruangsri2016} did not observe any resonances under the spin perturbation in their frequency-domain analysis, which they attributed to the existence of the R{\"u}diger constants (see section \ref{subsec:constint}). It seems plausible that the existence of the additional near-conserved quantities would suppress the resonances, but I will have to leave the question unanswered for a lack of conclusive arguments. 
 
 As for the question of chaos - for weakly perturbed integrable systems it is well known to occur only in a thin layer of the topological transition between the bulk of the motion and the resonant layer. The thickness of the chaotic layer caused by a smooth Hamiltonian perturbation of size $\varepsilon$ is even conjectured to scale as $\sim \exp(-\lambda/ |\varepsilon|)$ as $\varepsilon \to 0$ \cite{arnold2007}. Since no such topological transitions are even present in the foliation of phase space at given order, we can safely that chaotic motion caused by linear-in-spin perturbations is negligible. 
 
{\em Generalizations to other space-times.} The construction given in this paper has some obvious mathematical generalizations. It is possible to repeat the separation in an identical manner in the entire class of four-dimensional Kerr-NUT-(A)dS space-times \citep{carter1968b} and for massless spinning particles possibly even in the entire Pleba{\'n}ski-Demia{\'n}ski class of space-times \citep{plebanski1976}. Similarly, it seems to be easy to generalize the construction to Kerr-NUT-(A)dS space-times of dimension 5 by using the tetrad found by \citet{connell2008}. However, it has to be explicitly verified whether the growing number of degrees of freedom of a classical rotating body in growing dimension match or outpace the number of integrals of motion provided by the hidden symmetry (see \citep{kubiznak2012} for a discussion of this issue for a {\em semi-classical} spinning particle). 

{\em Implications for self-forced inspirals.} The computation of the shift to fundamental frequencies the way it is presented here provides almost all the necessary ingredients for the implementation of the conservative spin-curvature coupling into EMRI codes based on the two-timescale approximation scheme \citep{hinderer2008,vandemeent2018}. The only issue to resolve is a mapping of the spin-perturbed orbits to a set of fiducial geodesic that remains $\mathcal{O}(s)$ close to identity even for near-circular and circular orbits.

However, as already discussed in the Introduction, the non-negligible finite-size effects in EMRIs also include the dissipative decay of spin and a correction to the dissipation rates of the orbital constants of motion due to the spin perturbation to the trajectory. 

The two constants of motion in the spin sector are the aligned component of spin $s_\parallel = C_{\rm Y}/\sqrt{K}$ and the total spin magnitude $s$. The evolution of the spin tensor can be viewed as parallel transport in a certain smooth metric $g_{\mu\nu} + h_{\mu\nu}^{\rm R}$ \citep{detweiler2003,harte2012}. Since the magnitude of spin is conserved in any metric, we may deduce the immediate magnitude of spin $s_{ g}^2 \equiv s^{\mu\nu}s^{\kappa \lambda} g_{\mu\kappa}g_{\nu \lambda}/2$ from the conserved value of $s_{h+g}^2 = s_{g}^2 + s^{\mu\nu} s^{\kappa\lambda}g_{\mu\kappa}h^{\rm R}_{\nu\lambda}$ and the local value of the metric perturbation $h_{\mu\nu}^{\rm R}$.  The only non-trivial computation for the spin dissipation then reads
\begin{align}
    \begin{split}
        \langle \dot{C}_{\rm Y} \rangle_{\rm g} =& \left\langle \frac{1}{2} Y_{\mu\nu}  \varepsilon^{\mu\kappa\lambda\gamma} \left(f^\nu u_\kappa s_{\lambda \gamma} + u^\nu f_\kappa s_{\lambda \gamma} \right) \right\rangle_{\!\!\rm g} 
        \\ 
        &+ \left\langle \frac{1}{2} Y_{\mu\nu}  \varepsilon^{\mu\kappa\lambda\gamma} u^\nu u_\kappa \tau_{\lambda \gamma}  \right\rangle_{\!\!\rm g}\,,
    \end{split}
    \\
    f^\nu \equiv& -\delta \Gamma^\mu_{\nu\kappa} u^\nu u^\kappa \,,
    \\
    \tau^{\mu\nu} \equiv & (\delta \Gamma^{\mu}_{\;\;\lambda\kappa} s^{\nu\lambda} - \delta \Gamma^{\nu}_{\;\;\lambda\kappa} s^{\mu\lambda} )u^\kappa\,,
    \\
    \delta \Gamma^\mu_{\;\;\nu\kappa} \equiv& \frac{1}{2}g^{\mu\lambda}(h^{\rm R}_{\lambda\nu,\kappa} + h^{\rm R}_{\lambda\kappa,\nu} - h^{\rm R}_{\kappa \nu,\lambda})\,,
\end{align}
where $f^\nu$ is the self-force\footnote{Here the four-velocity is normalized in the effective metric $g_{\mu\nu} + h^{\rm R}_{\mu\nu}$ and the self-force is thus not projected on the subspace orthogonal to $u^\mu$ (see \citep{detweiler2008,sago2008}).} on the particle centroid and $\tau_{\lambda \gamma}$ the self-torque \citep{harte2012}. It should then be easy to adapt mode-sum averaging methods used for monopole particles such as in Refs. \citep{mino2003,mino2005,hughes2005,sago2006} for the purpose of the $\langle \dot{C}_{\rm Y} \rangle_{\rm g}$ computation. 

On the other hand, the changes in the dissipation rates of the orbital parameters due to the spin perturbation of the orbit requires the ability to compute averages $\langle \rangle_{\rm sp}$ of various functions over the spin-perturbed orbit. However, only averages of functions that are additively separable can be given in terms of simple geodesic averages such as in equation \eqref{eq:spavg}. In constrast, the mode-sum method of computing the metric perturbations naturally works with {\em multiplicatively} separable functions. Hence, it seems that one will need to compute a sufficient number of Fourier coefficients of $\delta f_y/f_y$ in order to construct the shift vector $\vec{\xi}$ (from equation \eqref{eq:xisol}) and use it in the $\langle \rangle_{\rm sp}$ averages. However, this will also require a careful treatment of the turning-point singularities that arise in the formalism.

\section*{Acknowledgments}
I would like to thank Jan Steinhoff, Justin Vines, Maarten van de Meent, and Georgios Lukes-Gerakopoulos for feedback and numerous discussions of the subject. Preliminary versions of the results in this paper were published in my Ph.D. thesis, which was kindly supervised by Claus L{\"a}mmerzahl and supported from a Ph.D. grant of the German Research Foundation (DFG) within Research Training Group 1620 ``Models of Gravity''. I am also supported by Grant No. GACR-17-06962Y of the Czech Science Foundation, which is acknowledged gratefully. 

\bibliography{literatura}

\begin{thebibliography}{89}%
\makeatletter
\providecommand \@ifxundefined [1]{%
 \@ifx{#1\undefined}
}%
\providecommand \@ifnum [1]{%
 \ifnum #1\expandafter \@firstoftwo
 \else \expandafter \@secondoftwo
 \fi
}%
\providecommand \@ifx [1]{%
 \ifx #1\expandafter \@firstoftwo
 \else \expandafter \@secondoftwo
 \fi
}%
\providecommand \natexlab [1]{#1}%
\providecommand \enquote  [1]{``#1''}%
\providecommand \bibnamefont  [1]{#1}%
\providecommand \bibfnamefont [1]{#1}%
\providecommand \citenamefont [1]{#1}%
\providecommand \href@noop [0]{\@secondoftwo}%
\providecommand \href [0]{\begingroup \@sanitize@url \@href}%
\providecommand \@href[1]{\@@startlink{#1}\@@href}%
\providecommand \@@href[1]{\endgroup#1\@@endlink}%
\providecommand \@sanitize@url [0]{\catcode `\\12\catcode `\$12\catcode
  `\&12\catcode `\#12\catcode `\^12\catcode `\_12\catcode `\%12\relax}%
\providecommand \@@startlink[1]{}%
\providecommand \@@endlink[0]{}%
\providecommand \url  [0]{\begingroup\@sanitize@url \@url }%
\providecommand \@url [1]{\endgroup\@href {#1}{\urlprefix }}%
\providecommand \urlprefix  [0]{URL }%
\providecommand \Eprint [0]{\href }%
\providecommand \doibase [0]{http://dx.doi.org/}%
\providecommand \selectlanguage [0]{\@gobble}%
\providecommand \bibinfo  [0]{\@secondoftwo}%
\providecommand \bibfield  [0]{\@secondoftwo}%
\providecommand \translation [1]{[#1]}%
\providecommand \BibitemOpen [0]{}%
\providecommand \bibitemStop [0]{}%
\providecommand \bibitemNoStop [0]{.\EOS\space}%
\providecommand \EOS [0]{\spacefactor3000\relax}%
\providecommand \BibitemShut  [1]{\csname bibitem#1\endcsname}%
\let\auto@bib@innerbib\@empty
\bibitem [{\citenamefont {Poisson}\ \emph {et~al.}(2011)\citenamefont
  {Poisson}, \citenamefont {Pound},\ and\ \citenamefont {Vega}}]{poisson2011}%
  \BibitemOpen
  \bibfield  {author} {\bibinfo {author} {\bibfnamefont {E.}~\bibnamefont
  {Poisson}}, \bibinfo {author} {\bibfnamefont {A.}~\bibnamefont {Pound}}, \
  and\ \bibinfo {author} {\bibfnamefont {I.}~\bibnamefont {Vega}},\ }\href
  {\doibase 10.12942/lrr-2011-7} {\bibfield  {journal} {\bibinfo  {journal}
  {Living Rev. Rel.}\ }\textbf {\bibinfo {volume} {14}},\ \bibinfo {pages} {7}
  (\bibinfo {year} {2011})},\ \Eprint {http://arxiv.org/abs/1102.0529}
  {arXiv:1102.0529 [gr-qc]} \BibitemShut {NoStop}%
\bibitem [{\citenamefont {Harte}(2012)}]{harte2012}%
  \BibitemOpen
  \bibfield  {author} {\bibinfo {author} {\bibfnamefont {A.~I.}\ \bibnamefont
  {Harte}},\ }\href@noop {} {\bibfield  {journal} {\bibinfo  {journal} {Class.
  Quant. Grav.}\ }\textbf {\bibinfo {volume} {29}},\ \bibinfo {pages} {055012}
  (\bibinfo {year} {2012})}\BibitemShut {NoStop}%
\bibitem [{\citenamefont {Barack}\ and\ \citenamefont
  {Pound}(2018)}]{barack2018}%
  \BibitemOpen
  \bibfield  {author} {\bibinfo {author} {\bibfnamefont {L.}~\bibnamefont
  {Barack}}\ and\ \bibinfo {author} {\bibfnamefont {A.}~\bibnamefont {Pound}},\
  }\href@noop {} {\bibfield  {journal} {\bibinfo  {journal} {Rep. Prog. Phys.}\
  }\textbf {\bibinfo {volume} {82}},\ \bibinfo {pages} {016904} (\bibinfo
  {year} {2018})},\ \Eprint {http://arxiv.org/abs/1805.10385} {arXiv:1805.10385
  [gr-qc]} \BibitemShut {NoStop}%
\bibitem [{\citenamefont {Gair}(2009)}]{gair2009}%
  \BibitemOpen
  \bibfield  {author} {\bibinfo {author} {\bibfnamefont {J.~R.}\ \bibnamefont
  {Gair}},\ }\href@noop {} {\bibfield  {journal} {\bibinfo  {journal} {Class.
  Quant. Grav.}\ }\textbf {\bibinfo {volume} {26}},\ \bibinfo {pages} {094034}
  (\bibinfo {year} {2009})}\BibitemShut {NoStop}%
\bibitem [{\citenamefont {{LISA Consortium}}(2017)}]{amaro2017}%
  \BibitemOpen
  \bibfield  {author} {\bibinfo {author} {\bibnamefont {{LISA Consortium}}},\
  }\href@noop {} {\  (\bibinfo {year} {2017})},\ \Eprint
  {http://arxiv.org/abs/1702.00786} {arXiv:1702.00786 [astro-ph]} \BibitemShut
  {NoStop}%
\bibitem [{\citenamefont {Hinderer}\ and\ \citenamefont
  {Flanagan}(2008)}]{hinderer2008}%
  \BibitemOpen
  \bibfield  {author} {\bibinfo {author} {\bibfnamefont {T.}~\bibnamefont
  {Hinderer}}\ and\ \bibinfo {author} {\bibfnamefont {E.~E.}\ \bibnamefont
  {Flanagan}},\ }\href {\doibase 10.1103/PhysRevD.78.064028} {\bibfield
  {journal} {\bibinfo  {journal} {Phys. Rev.}\ }\textbf {\bibinfo {volume}
  {D78}},\ \bibinfo {pages} {064028} (\bibinfo {year} {2008})},\ \Eprint
  {http://arxiv.org/abs/0805.3337} {arXiv:0805.3337 [gr-qc]} \BibitemShut
  {NoStop}%
\bibitem [{\citenamefont {Semer{\'a}k}(1999)}]{semerak1999}%
  \BibitemOpen
  \bibfield  {author} {\bibinfo {author} {\bibfnamefont {O.}~\bibnamefont
  {Semer{\'a}k}},\ }\href {\doibase 10.1046/j.1365-8711.1999.02754.x}
  {\bibfield  {journal} {\bibinfo  {journal} {Mon. Not. R. Astron. Soc.}\
  }\textbf {\bibinfo {volume} {308}},\ \bibinfo {pages} {863} (\bibinfo {year}
  {1999})}\BibitemShut {NoStop}%
\bibitem [{\citenamefont {Kyrian}\ and\ \citenamefont
  {Semer{\'a}k}(2007)}]{kyrsem}%
  \BibitemOpen
  \bibfield  {author} {\bibinfo {author} {\bibfnamefont {K.}~\bibnamefont
  {Kyrian}}\ and\ \bibinfo {author} {\bibfnamefont {O.}~\bibnamefont
  {Semer{\'a}k}},\ }\href {\doibase 10.1111/j.1365-2966.2007.12502.x}
  {\bibfield  {journal} {\bibinfo  {journal} {Mon. Not. Roy. Astron. Soc.}\
  }\textbf {\bibinfo {volume} {382}},\ \bibinfo {pages} {1922} (\bibinfo {year}
  {2007})}\BibitemShut {NoStop}%
\bibitem [{\citenamefont {Tanaka}\ \emph {et~al.}(1996)\citenamefont {Tanaka},
  \citenamefont {Mino}, \citenamefont {Sasaki},\ and\ \citenamefont
  {Shibata}}]{tanaka1996}%
  \BibitemOpen
  \bibfield  {author} {\bibinfo {author} {\bibfnamefont {T.}~\bibnamefont
  {Tanaka}}, \bibinfo {author} {\bibfnamefont {Y.}~\bibnamefont {Mino}},
  \bibinfo {author} {\bibfnamefont {M.}~\bibnamefont {Sasaki}}, \ and\ \bibinfo
  {author} {\bibfnamefont {M.}~\bibnamefont {Shibata}},\ }\href@noop {}
  {\bibfield  {journal} {\bibinfo  {journal} {Phys. Rev. D}\ }\textbf {\bibinfo
  {volume} {54}},\ \bibinfo {pages} {3762} (\bibinfo {year}
  {1996})}\BibitemShut {NoStop}%
\bibitem [{\citenamefont {Suzuki}\ and\ \citenamefont
  {Maeda}(1997)}]{suzuki1997}%
  \BibitemOpen
  \bibfield  {author} {\bibinfo {author} {\bibfnamefont {S.}~\bibnamefont
  {Suzuki}}\ and\ \bibinfo {author} {\bibfnamefont {K.-i.}\ \bibnamefont
  {Maeda}},\ }\href {\doibase 10.1103/PhysRevD.55.4848} {\bibfield  {journal}
  {\bibinfo  {journal} {Phys. Rev.}\ }\textbf {\bibinfo {volume} {D55}},\
  \bibinfo {pages} {4848} (\bibinfo {year} {1997})},\ \Eprint
  {http://arxiv.org/abs/gr-qc/9604020} {arXiv:gr-qc/9604020 [gr-qc]}
  \BibitemShut {NoStop}%
\bibitem [{\citenamefont {Tominaga}\ \emph {et~al.}(2001)\citenamefont
  {Tominaga}, \citenamefont {Saijo},\ and\ \citenamefont
  {Maeda}}]{tominaga2001}%
  \BibitemOpen
  \bibfield  {author} {\bibinfo {author} {\bibfnamefont {K.}~\bibnamefont
  {Tominaga}}, \bibinfo {author} {\bibfnamefont {M.}~\bibnamefont {Saijo}}, \
  and\ \bibinfo {author} {\bibfnamefont {K.-i.}\ \bibnamefont {Maeda}},\
  }\href@noop {} {\bibfield  {journal} {\bibinfo  {journal} {Phys. Rev. D}\
  }\textbf {\bibinfo {volume} {63}},\ \bibinfo {pages} {124012} (\bibinfo
  {year} {2001})}\BibitemShut {NoStop}%
\bibitem [{\citenamefont {{Harms}}\ \emph {et~al.}(2016)\citenamefont
  {{Harms}}, \citenamefont {{Lukes-Gerakopoulos}}, \citenamefont {{Bernuzzi}},\
  and\ \citenamefont {{Nagar}}}]{Harms2016}%
  \BibitemOpen
  \bibfield  {author} {\bibinfo {author} {\bibfnamefont {E.}~\bibnamefont
  {{Harms}}}, \bibinfo {author} {\bibfnamefont {G.}~\bibnamefont
  {{Lukes-Gerakopoulos}}}, \bibinfo {author} {\bibfnamefont {S.}~\bibnamefont
  {{Bernuzzi}}}, \ and\ \bibinfo {author} {\bibfnamefont {A.}~\bibnamefont
  {{Nagar}}},\ }\href {\doibase 10.1103/PhysRevD.94.104010} {\bibfield
  {journal} {\bibinfo  {journal} {Phys. Rev. D}\ }\textbf {\bibinfo {volume}
  {94}},\ \bibinfo {eid} {104010} (\bibinfo {year} {2016})},\ \Eprint
  {http://arxiv.org/abs/1609.00356} {arXiv:1609.00356 [gr-qc]} \BibitemShut
  {NoStop}%
\bibitem [{\citenamefont {Lukes-Gerakopoulos}\ \emph
  {et~al.}(2017)\citenamefont {Lukes-Gerakopoulos}, \citenamefont {Harms},
  \citenamefont {Bernuzzi},\ and\ \citenamefont {Nagar}}]{lukes2017}%
  \BibitemOpen
  \bibfield  {author} {\bibinfo {author} {\bibfnamefont {G.}~\bibnamefont
  {Lukes-Gerakopoulos}}, \bibinfo {author} {\bibfnamefont {E.}~\bibnamefont
  {Harms}}, \bibinfo {author} {\bibfnamefont {S.}~\bibnamefont {Bernuzzi}}, \
  and\ \bibinfo {author} {\bibfnamefont {A.}~\bibnamefont {Nagar}},\
  }\href@noop {} {\bibfield  {journal} {\bibinfo  {journal} {Phys. Rev. D}\
  }\textbf {\bibinfo {volume} {96}},\ \bibinfo {pages} {064051} (\bibinfo
  {year} {2017})}\BibitemShut {NoStop}%
\bibitem [{\citenamefont {Dolan}\ \emph {et~al.}(2014)\citenamefont {Dolan},
  \citenamefont {Warburton}, \citenamefont {Harte}, \citenamefont {Le~Tiec},
  \citenamefont {Wardell},\ and\ \citenamefont {Barack}}]{dolan2014}%
  \BibitemOpen
  \bibfield  {author} {\bibinfo {author} {\bibfnamefont {S.~R.}\ \bibnamefont
  {Dolan}}, \bibinfo {author} {\bibfnamefont {N.}~\bibnamefont {Warburton}},
  \bibinfo {author} {\bibfnamefont {A.~I.}\ \bibnamefont {Harte}}, \bibinfo
  {author} {\bibfnamefont {A.}~\bibnamefont {Le~Tiec}}, \bibinfo {author}
  {\bibfnamefont {B.}~\bibnamefont {Wardell}}, \ and\ \bibinfo {author}
  {\bibfnamefont {L.}~\bibnamefont {Barack}},\ }\href@noop {} {\bibfield
  {journal} {\bibinfo  {journal} {Phys. Rev. D}\ }\textbf {\bibinfo {volume}
  {89}},\ \bibinfo {pages} {064011} (\bibinfo {year} {2014})}\BibitemShut
  {NoStop}%
\bibitem [{\citenamefont {Akcay}\ \emph {et~al.}(2017)\citenamefont {Akcay},
  \citenamefont {Dempsey},\ and\ \citenamefont {Dolan}}]{akcay2017}%
  \BibitemOpen
  \bibfield  {author} {\bibinfo {author} {\bibfnamefont {S.}~\bibnamefont
  {Akcay}}, \bibinfo {author} {\bibfnamefont {D.}~\bibnamefont {Dempsey}}, \
  and\ \bibinfo {author} {\bibfnamefont {S.~R.}\ \bibnamefont {Dolan}},\
  }\href@noop {} {\bibfield  {journal} {\bibinfo  {journal} {Class. Quant.
  Grav.}\ }\textbf {\bibinfo {volume} {34}},\ \bibinfo {pages} {084001}
  (\bibinfo {year} {2017})}\BibitemShut {NoStop}%
\bibitem [{\citenamefont {Kavanagh}\ \emph {et~al.}(2017)\citenamefont
  {Kavanagh}, \citenamefont {Bini}, \citenamefont {Damour}, \citenamefont
  {Hopper}, \citenamefont {Ottewill},\ and\ \citenamefont
  {Wardell}}]{kavanagh2017}%
  \BibitemOpen
  \bibfield  {author} {\bibinfo {author} {\bibfnamefont {C.}~\bibnamefont
  {Kavanagh}}, \bibinfo {author} {\bibfnamefont {D.}~\bibnamefont {Bini}},
  \bibinfo {author} {\bibfnamefont {T.}~\bibnamefont {Damour}}, \bibinfo
  {author} {\bibfnamefont {S.}~\bibnamefont {Hopper}}, \bibinfo {author}
  {\bibfnamefont {A.~C.}\ \bibnamefont {Ottewill}}, \ and\ \bibinfo {author}
  {\bibfnamefont {B.}~\bibnamefont {Wardell}},\ }\href {\doibase
  10.1103/PhysRevD.96.064012} {\bibfield  {journal} {\bibinfo  {journal} {Phys.
  Rev.}\ }\textbf {\bibinfo {volume} {D96}},\ \bibinfo {pages} {064012}
  (\bibinfo {year} {2017})},\ \Eprint {http://arxiv.org/abs/1706.00459}
  {arXiv:1706.00459 [gr-qc]} \BibitemShut {NoStop}%
\bibitem [{\citenamefont {Akcay}(2017)}]{akcay2017b}%
  \BibitemOpen
  \bibfield  {author} {\bibinfo {author} {\bibfnamefont {S.}~\bibnamefont
  {Akcay}},\ }\href@noop {} {\bibfield  {journal} {\bibinfo  {journal} {Phys.
  Rev. D}\ }\textbf {\bibinfo {volume} {96}},\ \bibinfo {pages} {044024}
  (\bibinfo {year} {2017})}\BibitemShut {NoStop}%
\bibitem [{\citenamefont {Bini}\ \emph {et~al.}(2018)\citenamefont {Bini},
  \citenamefont {Damour}, \citenamefont {Geralico}, \citenamefont {Kavanagh},\
  and\ \citenamefont {Van~de Meent}}]{bini2018}%
  \BibitemOpen
  \bibfield  {author} {\bibinfo {author} {\bibfnamefont {D.}~\bibnamefont
  {Bini}}, \bibinfo {author} {\bibfnamefont {T.}~\bibnamefont {Damour}},
  \bibinfo {author} {\bibfnamefont {A.}~\bibnamefont {Geralico}}, \bibinfo
  {author} {\bibfnamefont {C.}~\bibnamefont {Kavanagh}}, \ and\ \bibinfo
  {author} {\bibfnamefont {M.}~\bibnamefont {Van~de Meent}},\ }\href {\doibase
  10.1103/physrevd.98.104062} {\bibfield  {journal} {\bibinfo  {journal} {Phys.
  Rev. D}\ }\textbf {\bibinfo {volume} {98}},\ \bibinfo {pages} {104062}
  (\bibinfo {year} {2018})}\BibitemShut {NoStop}%
\bibitem [{\citenamefont {Mashhoon}\ and\ \citenamefont
  {Singh}(2006)}]{mashhoon2006}%
  \BibitemOpen
  \bibfield  {author} {\bibinfo {author} {\bibfnamefont {B.}~\bibnamefont
  {Mashhoon}}\ and\ \bibinfo {author} {\bibfnamefont {D.}~\bibnamefont
  {Singh}},\ }\href@noop {} {\bibfield  {journal} {\bibinfo  {journal} {Phys.
  Rev. D}\ }\textbf {\bibinfo {volume} {74}},\ \bibinfo {pages} {124006}
  (\bibinfo {year} {2006})}\BibitemShut {NoStop}%
\bibitem [{\citenamefont {Singh}(2008)}]{singh2008}%
  \BibitemOpen
  \bibfield  {author} {\bibinfo {author} {\bibfnamefont {D.}~\bibnamefont
  {Singh}},\ }\href@noop {} {\bibfield  {journal} {\bibinfo  {journal} {Phys.
  Rev. D}\ }\textbf {\bibinfo {volume} {78}},\ \bibinfo {pages} {104028}
  (\bibinfo {year} {2008})}\BibitemShut {NoStop}%
\bibitem [{\citenamefont {Bini}\ \emph {et~al.}(2011)\citenamefont {Bini},
  \citenamefont {Geralico},\ and\ \citenamefont {Jantzen}}]{bini2011}%
  \BibitemOpen
  \bibfield  {author} {\bibinfo {author} {\bibfnamefont {D.}~\bibnamefont
  {Bini}}, \bibinfo {author} {\bibfnamefont {A.}~\bibnamefont {Geralico}}, \
  and\ \bibinfo {author} {\bibfnamefont {R.~T.}\ \bibnamefont {Jantzen}},\
  }\href@noop {} {\bibfield  {journal} {\bibinfo  {journal} {Gen. Relativ.
  Gravitation}\ }\textbf {\bibinfo {volume} {43}},\ \bibinfo {pages} {959}
  (\bibinfo {year} {2011})}\BibitemShut {NoStop}%
\bibitem [{\citenamefont {Bini}\ and\ \citenamefont
  {Geralico}(2011)}]{bini2011a}%
  \BibitemOpen
  \bibfield  {author} {\bibinfo {author} {\bibfnamefont {D.}~\bibnamefont
  {Bini}}\ and\ \bibinfo {author} {\bibfnamefont {A.}~\bibnamefont
  {Geralico}},\ }\href@noop {} {\bibfield  {journal} {\bibinfo  {journal}
  {Phys. Rev. D}\ }\textbf {\bibinfo {volume} {84}},\ \bibinfo {pages} {104012}
  (\bibinfo {year} {2011})}\BibitemShut {NoStop}%
\bibitem [{\citenamefont {Huerta}\ and\ \citenamefont
  {Gair}(2011)}]{huerta2011}%
  \BibitemOpen
  \bibfield  {author} {\bibinfo {author} {\bibfnamefont {E.}~\bibnamefont
  {Huerta}}\ and\ \bibinfo {author} {\bibfnamefont {J.~R.}\ \bibnamefont
  {Gair}},\ }\href@noop {} {\bibfield  {journal} {\bibinfo  {journal} {Phys.
  Rev. D}\ }\textbf {\bibinfo {volume} {84}},\ \bibinfo {pages} {064023}
  (\bibinfo {year} {2011})}\BibitemShut {NoStop}%
\bibitem [{\citenamefont {Huerta}\ \emph {et~al.}(2012)\citenamefont {Huerta},
  \citenamefont {Gair},\ and\ \citenamefont {Brown}}]{huerta2012}%
  \BibitemOpen
  \bibfield  {author} {\bibinfo {author} {\bibfnamefont {E.}~\bibnamefont
  {Huerta}}, \bibinfo {author} {\bibfnamefont {J.~R.}\ \bibnamefont {Gair}}, \
  and\ \bibinfo {author} {\bibfnamefont {D.~A.}\ \bibnamefont {Brown}},\
  }\href@noop {} {\bibfield  {journal} {\bibinfo  {journal} {Phys. Rev. D}\
  }\textbf {\bibinfo {volume} {85}},\ \bibinfo {pages} {064023} (\bibinfo
  {year} {2012})}\BibitemShut {NoStop}%
\bibitem [{\citenamefont {Burko}\ and\ \citenamefont
  {Khanna}(2015)}]{burko2015}%
  \BibitemOpen
  \bibfield  {author} {\bibinfo {author} {\bibfnamefont {L.~M.}\ \bibnamefont
  {Burko}}\ and\ \bibinfo {author} {\bibfnamefont {G.}~\bibnamefont {Khanna}},\
  }\href@noop {} {\bibfield  {journal} {\bibinfo  {journal} {Phys. Rev. D}\
  }\textbf {\bibinfo {volume} {91}},\ \bibinfo {pages} {104017} (\bibinfo
  {year} {2015})}\BibitemShut {NoStop}%
\bibitem [{\citenamefont {Warburton}\ \emph {et~al.}(2017)\citenamefont
  {Warburton}, \citenamefont {Osburn},\ and\ \citenamefont
  {Evans}}]{Warburton2017}%
  \BibitemOpen
  \bibfield  {author} {\bibinfo {author} {\bibfnamefont {N.}~\bibnamefont
  {Warburton}}, \bibinfo {author} {\bibfnamefont {T.}~\bibnamefont {Osburn}}, \
  and\ \bibinfo {author} {\bibfnamefont {C.~R.}\ \bibnamefont {Evans}},\ }\href
  {\doibase 10.1103/PhysRevD.96.084057} {\bibfield  {journal} {\bibinfo
  {journal} {Phys. Rev.}\ }\textbf {\bibinfo {volume} {D96}},\ \bibinfo {pages}
  {084057} (\bibinfo {year} {2017})},\ \Eprint
  {http://arxiv.org/abs/1708.03720} {arXiv:1708.03720 [gr-qc]} \BibitemShut
  {NoStop}%
\bibitem [{\citenamefont {Will}\ and\ \citenamefont {Maitra}(2017)}]{will2017}%
  \BibitemOpen
  \bibfield  {author} {\bibinfo {author} {\bibfnamefont {C.~M.}\ \bibnamefont
  {Will}}\ and\ \bibinfo {author} {\bibfnamefont {M.}~\bibnamefont {Maitra}},\
  }\href@noop {} {\bibfield  {journal} {\bibinfo  {journal} {Phys. Rev. D}\
  }\textbf {\bibinfo {volume} {95}},\ \bibinfo {pages} {064003} (\bibinfo
  {year} {2017})}\BibitemShut {NoStop}%
\bibitem [{\citenamefont {Arnold}\ \emph {et~al.}(2007)\citenamefont {Arnold},
  \citenamefont {Kozlov},\ and\ \citenamefont {Neishtadt}}]{arnold2007}%
  \BibitemOpen
  \bibfield  {author} {\bibinfo {author} {\bibfnamefont {V.~I.}\ \bibnamefont
  {Arnold}}, \bibinfo {author} {\bibfnamefont {V.~V.}\ \bibnamefont {Kozlov}},
  \ and\ \bibinfo {author} {\bibfnamefont {A.~I.}\ \bibnamefont {Neishtadt}},\
  }\href@noop {} {\emph {\bibinfo {title} {Mathematical aspects of classical
  and celestial mechanics}}},\ Vol.~\bibinfo {volume} {3}\ (\bibinfo
  {publisher} {Springer Science \& Business Media},\ \bibinfo {year}
  {2007})\BibitemShut {NoStop}%
\bibitem [{\citenamefont {Flanagan}\ and\ \citenamefont
  {Hinderer}(2012)}]{flanagan2012}%
  \BibitemOpen
  \bibfield  {author} {\bibinfo {author} {\bibfnamefont {E.~E.}\ \bibnamefont
  {Flanagan}}\ and\ \bibinfo {author} {\bibfnamefont {T.}~\bibnamefont
  {Hinderer}},\ }\href@noop {} {\bibfield  {journal} {\bibinfo  {journal}
  {Phys. Rev. Lett.}\ }\textbf {\bibinfo {volume} {109}},\ \bibinfo {pages}
  {071102} (\bibinfo {year} {2012})}\BibitemShut {NoStop}%
\bibitem [{\citenamefont {Brink}\ \emph
  {et~al.}(2015{\natexlab{a}})\citenamefont {Brink}, \citenamefont {Geyer},\
  and\ \citenamefont {Hinderer}}]{brink2015}%
  \BibitemOpen
  \bibfield  {author} {\bibinfo {author} {\bibfnamefont {J.}~\bibnamefont
  {Brink}}, \bibinfo {author} {\bibfnamefont {M.}~\bibnamefont {Geyer}}, \ and\
  \bibinfo {author} {\bibfnamefont {T.}~\bibnamefont {Hinderer}},\ }\href@noop
  {} {\bibfield  {journal} {\bibinfo  {journal} {Phys. Rev. Lett.}\ }\textbf
  {\bibinfo {volume} {114}},\ \bibinfo {pages} {081102} (\bibinfo {year}
  {2015}{\natexlab{a}})}\BibitemShut {NoStop}%
\bibitem [{\citenamefont {Brink}\ \emph
  {et~al.}(2015{\natexlab{b}})\citenamefont {Brink}, \citenamefont {Geyer},\
  and\ \citenamefont {Hinderer}}]{brink2015b}%
  \BibitemOpen
  \bibfield  {author} {\bibinfo {author} {\bibfnamefont {J.}~\bibnamefont
  {Brink}}, \bibinfo {author} {\bibfnamefont {M.}~\bibnamefont {Geyer}}, \ and\
  \bibinfo {author} {\bibfnamefont {T.}~\bibnamefont {Hinderer}},\ }\href@noop
  {} {\bibfield  {journal} {\bibinfo  {journal} {Phys. Rev. D}\ }\textbf
  {\bibinfo {volume} {91}},\ \bibinfo {pages} {083001} (\bibinfo {year}
  {2015}{\natexlab{b}})}\BibitemShut {NoStop}%
\bibitem [{\citenamefont {Flanagan}\ \emph {et~al.}(2014)\citenamefont
  {Flanagan}, \citenamefont {Hughes},\ and\ \citenamefont
  {Ruangsri}}]{flanagan2014}%
  \BibitemOpen
  \bibfield  {author} {\bibinfo {author} {\bibfnamefont {E.~E.}\ \bibnamefont
  {Flanagan}}, \bibinfo {author} {\bibfnamefont {S.~A.}\ \bibnamefont
  {Hughes}}, \ and\ \bibinfo {author} {\bibfnamefont {U.}~\bibnamefont
  {Ruangsri}},\ }\href@noop {} {\bibfield  {journal} {\bibinfo  {journal}
  {Phys. Rev. D}\ }\textbf {\bibinfo {volume} {89}},\ \bibinfo {pages} {084028}
  (\bibinfo {year} {2014})}\BibitemShut {NoStop}%
\bibitem [{\citenamefont {van~de Meent}(2014{\natexlab{a}})}]{vandemeent2014}%
  \BibitemOpen
  \bibfield  {author} {\bibinfo {author} {\bibfnamefont {M.}~\bibnamefont
  {van~de Meent}},\ }\href@noop {} {\bibfield  {journal} {\bibinfo  {journal}
  {Phys. Rev. D}\ }\textbf {\bibinfo {volume} {90}},\ \bibinfo {pages} {044027}
  (\bibinfo {year} {2014}{\natexlab{a}})}\BibitemShut {NoStop}%
\bibitem [{\citenamefont {van~de Meent}(2014{\natexlab{b}})}]{vandemeent2014b}%
  \BibitemOpen
  \bibfield  {author} {\bibinfo {author} {\bibfnamefont {M.}~\bibnamefont
  {van~de Meent}},\ }\href@noop {} {\bibfield  {journal} {\bibinfo  {journal}
  {Phys. Rev. D}\ }\textbf {\bibinfo {volume} {89}},\ \bibinfo {pages} {084033}
  (\bibinfo {year} {2014}{\natexlab{b}})}\BibitemShut {NoStop}%
\bibitem [{\citenamefont {Lewis}\ \emph {et~al.}(2017)\citenamefont {Lewis},
  \citenamefont {Zimmerman},\ and\ \citenamefont {Pfeiffer}}]{lewis2017}%
  \BibitemOpen
  \bibfield  {author} {\bibinfo {author} {\bibfnamefont {A.~G.}\ \bibnamefont
  {Lewis}}, \bibinfo {author} {\bibfnamefont {A.}~\bibnamefont {Zimmerman}}, \
  and\ \bibinfo {author} {\bibfnamefont {H.~P.}\ \bibnamefont {Pfeiffer}},\
  }\href@noop {} {\bibfield  {journal} {\bibinfo  {journal} {Class. Quant.
  Grav.}\ }\textbf {\bibinfo {volume} {34}},\ \bibinfo {pages} {124001}
  (\bibinfo {year} {2017})}\BibitemShut {NoStop}%
\bibitem [{\citenamefont {Apostolatos}\ \emph {et~al.}(2009)\citenamefont
  {Apostolatos}, \citenamefont {Lukes-Gerakopoulos},\ and\ \citenamefont
  {Contopoulos}}]{apostolatos2009}%
  \BibitemOpen
  \bibfield  {author} {\bibinfo {author} {\bibfnamefont {T.~A.}\ \bibnamefont
  {Apostolatos}}, \bibinfo {author} {\bibfnamefont {G.}~\bibnamefont
  {Lukes-Gerakopoulos}}, \ and\ \bibinfo {author} {\bibfnamefont
  {G.}~\bibnamefont {Contopoulos}},\ }\href@noop {} {\bibfield  {journal}
  {\bibinfo  {journal} {Phys. Rev. Lett.}\ }\textbf {\bibinfo {volume} {103}},\
  \bibinfo {pages} {111101} (\bibinfo {year} {2009})}\BibitemShut {NoStop}%
\bibitem [{\citenamefont {Lukes-Gerakopoulos}\ \emph
  {et~al.}(2010)\citenamefont {Lukes-Gerakopoulos}, \citenamefont
  {Apostolatos},\ and\ \citenamefont {Contopoulos}}]{lukes2010}%
  \BibitemOpen
  \bibfield  {author} {\bibinfo {author} {\bibfnamefont {G.}~\bibnamefont
  {Lukes-Gerakopoulos}}, \bibinfo {author} {\bibfnamefont {T.~A.}\ \bibnamefont
  {Apostolatos}}, \ and\ \bibinfo {author} {\bibfnamefont {G.}~\bibnamefont
  {Contopoulos}},\ }\href@noop {} {\bibfield  {journal} {\bibinfo  {journal}
  {Phys. Rev. D}\ }\textbf {\bibinfo {volume} {81}},\ \bibinfo {pages} {124005}
  (\bibinfo {year} {2010})}\BibitemShut {NoStop}%
\bibitem [{\citenamefont {Ruangsri}\ and\ \citenamefont
  {Hughes}(2014)}]{ruangsri2014}%
  \BibitemOpen
  \bibfield  {author} {\bibinfo {author} {\bibfnamefont {U.}~\bibnamefont
  {Ruangsri}}\ and\ \bibinfo {author} {\bibfnamefont {S.~A.}\ \bibnamefont
  {Hughes}},\ }\href@noop {} {\bibfield  {journal} {\bibinfo  {journal} {Phys.
  Rev. D}\ }\textbf {\bibinfo {volume} {89}},\ \bibinfo {pages} {084036}
  (\bibinfo {year} {2014})}\BibitemShut {NoStop}%
\bibitem [{\citenamefont {Berry}\ \emph {et~al.}(2016)\citenamefont {Berry},
  \citenamefont {Cole}, \citenamefont {Ca{\~n}izares},\ and\ \citenamefont
  {Gair}}]{berry2016}%
  \BibitemOpen
  \bibfield  {author} {\bibinfo {author} {\bibfnamefont {C.~P.}\ \bibnamefont
  {Berry}}, \bibinfo {author} {\bibfnamefont {R.~H.}\ \bibnamefont {Cole}},
  \bibinfo {author} {\bibfnamefont {P.}~\bibnamefont {Ca{\~n}izares}}, \ and\
  \bibinfo {author} {\bibfnamefont {J.~R.}\ \bibnamefont {Gair}},\ }\href@noop
  {} {\bibfield  {journal} {\bibinfo  {journal} {Phys. Rev. D}\ }\textbf
  {\bibinfo {volume} {94}},\ \bibinfo {pages} {124042} (\bibinfo {year}
  {2016})}\BibitemShut {NoStop}%
\bibitem [{\citenamefont {R{\"u}diger}(1981)}]{ruediger1}%
  \BibitemOpen
  \bibfield  {author} {\bibinfo {author} {\bibfnamefont {R.}~\bibnamefont
  {R{\"u}diger}},\ }\href@noop {} {\bibfield  {journal} {\bibinfo  {journal}
  {Proc. Royal Soc. Lond. A}\ }\textbf {\bibinfo {volume} {375}},\ \bibinfo
  {pages} {185} (\bibinfo {year} {1981})}\BibitemShut {NoStop}%
\bibitem [{\citenamefont {R{\"u}diger}(1983)}]{ruediger2}%
  \BibitemOpen
  \bibfield  {author} {\bibinfo {author} {\bibfnamefont {R.}~\bibnamefont
  {R{\"u}diger}},\ }\href@noop {} {\bibfield  {journal} {\bibinfo  {journal}
  {Proc. Royal Soc. Lond. A}\ }\textbf {\bibinfo {volume} {385}},\ \bibinfo
  {pages} {229} (\bibinfo {year} {1983})}\BibitemShut {NoStop}%
\bibitem [{\citenamefont {Apostolatos}(1996)}]{apostolatos1996}%
  \BibitemOpen
  \bibfield  {author} {\bibinfo {author} {\bibfnamefont {T.~A.}\ \bibnamefont
  {Apostolatos}},\ }\href@noop {} {\bibfield  {journal} {\bibinfo  {journal}
  {Class. Quant. Grav.}\ }\textbf {\bibinfo {volume} {13}},\ \bibinfo {pages}
  {799} (\bibinfo {year} {1996})}\BibitemShut {NoStop}%
\bibitem [{\citenamefont {Kunst}\ \emph {et~al.}(2016)\citenamefont {Kunst},
  \citenamefont {Ledvinka}, \citenamefont {Lukes-Gerakopoulos},\ and\
  \citenamefont {Seyrich}}]{kunst2016}%
  \BibitemOpen
  \bibfield  {author} {\bibinfo {author} {\bibfnamefont {D.}~\bibnamefont
  {Kunst}}, \bibinfo {author} {\bibfnamefont {T.}~\bibnamefont {Ledvinka}},
  \bibinfo {author} {\bibfnamefont {G.}~\bibnamefont {Lukes-Gerakopoulos}}, \
  and\ \bibinfo {author} {\bibfnamefont {J.}~\bibnamefont {Seyrich}},\ }\href
  {\doibase 10.1103/PhysRevD.93.044004} {\bibfield  {journal} {\bibinfo
  {journal} {Phys. Rev.}\ }\textbf {\bibinfo {volume} {D93}},\ \bibinfo {pages}
  {044004} (\bibinfo {year} {2016})},\ \Eprint
  {http://arxiv.org/abs/1506.01473} {arXiv:1506.01473 [gr-qc]} \BibitemShut
  {NoStop}%
\bibitem [{\citenamefont {Gibbons}\ \emph {et~al.}(1993)\citenamefont
  {Gibbons}, \citenamefont {Rietdijk},\ and\ \citenamefont
  {Van~Holten}}]{gibbons1993}%
  \BibitemOpen
  \bibfield  {author} {\bibinfo {author} {\bibfnamefont {G.}~\bibnamefont
  {Gibbons}}, \bibinfo {author} {\bibfnamefont {R.~H.}\ \bibnamefont
  {Rietdijk}}, \ and\ \bibinfo {author} {\bibfnamefont {J.}~\bibnamefont
  {Van~Holten}},\ }\href@noop {} {\bibfield  {journal} {\bibinfo  {journal}
  {Nucl. Phys. B}\ }\textbf {\bibinfo {volume} {404}},\ \bibinfo {pages} {42}
  (\bibinfo {year} {1993})}\BibitemShut {NoStop}%
\bibitem [{\citenamefont {Tanimoto}(1995)}]{tanimoto1995}%
  \BibitemOpen
  \bibfield  {author} {\bibinfo {author} {\bibfnamefont {M.}~\bibnamefont
  {Tanimoto}},\ }\href@noop {} {\bibfield  {journal} {\bibinfo  {journal}
  {Nucl. Phys. B}\ }\textbf {\bibinfo {volume} {442}},\ \bibinfo {pages} {549}
  (\bibinfo {year} {1995})}\BibitemShut {NoStop}%
\bibitem [{\citenamefont {Ahmedov}\ and\ \citenamefont
  {Aliev}(2009)}]{ahmedov2009}%
  \BibitemOpen
  \bibfield  {author} {\bibinfo {author} {\bibfnamefont {H.}~\bibnamefont
  {Ahmedov}}\ and\ \bibinfo {author} {\bibfnamefont {A.~N.}\ \bibnamefont
  {Aliev}},\ }\href@noop {} {\bibfield  {journal} {\bibinfo  {journal} {Phys.
  Rev. D}\ }\textbf {\bibinfo {volume} {79}},\ \bibinfo {pages} {084019}
  (\bibinfo {year} {2009})}\BibitemShut {NoStop}%
\bibitem [{\citenamefont {Kubiz{\v{n}}{\'a}k}\ and\ \citenamefont
  {Cariglia}(2012)}]{kubiznak2012}%
  \BibitemOpen
  \bibfield  {author} {\bibinfo {author} {\bibfnamefont {D.}~\bibnamefont
  {Kubiz{\v{n}}{\'a}k}}\ and\ \bibinfo {author} {\bibfnamefont
  {M.}~\bibnamefont {Cariglia}},\ }\href@noop {} {\bibfield  {journal}
  {\bibinfo  {journal} {Phys. Rev. Lett.}\ }\textbf {\bibinfo {volume} {108}},\
  \bibinfo {pages} {051104} (\bibinfo {year} {2012})}\BibitemShut {NoStop}%
\bibitem [{\citenamefont {Ruangsri}\ \emph {et~al.}(2016)\citenamefont
  {Ruangsri}, \citenamefont {Vigeland},\ and\ \citenamefont
  {Hughes}}]{ruangsri2016}%
  \BibitemOpen
  \bibfield  {author} {\bibinfo {author} {\bibfnamefont {U.}~\bibnamefont
  {Ruangsri}}, \bibinfo {author} {\bibfnamefont {S.~J.}\ \bibnamefont
  {Vigeland}}, \ and\ \bibinfo {author} {\bibfnamefont {S.~A.}\ \bibnamefont
  {Hughes}},\ }\href@noop {} {\bibfield  {journal} {\bibinfo  {journal} {Phys.
  Rev. D}\ }\textbf {\bibinfo {volume} {94}},\ \bibinfo {pages} {044008}
  (\bibinfo {year} {2016})}\BibitemShut {NoStop}%
\bibitem [{\citenamefont {Boyer}\ and\ \citenamefont
  {Lindquist}(1967)}]{boyer1967}%
  \BibitemOpen
  \bibfield  {author} {\bibinfo {author} {\bibfnamefont {R.~H.}\ \bibnamefont
  {Boyer}}\ and\ \bibinfo {author} {\bibfnamefont {R.~W.}\ \bibnamefont
  {Lindquist}},\ }\href@noop {} {\bibfield  {journal} {\bibinfo  {journal} {J.
  Math. Phys.}\ }\textbf {\bibinfo {volume} {8}},\ \bibinfo {pages} {265}
  (\bibinfo {year} {1967})}\BibitemShut {NoStop}%
\bibitem [{\citenamefont {Floyd}(1973)}]{floyd1973}%
  \BibitemOpen
  \bibfield  {author} {\bibinfo {author} {\bibfnamefont {R.}~\bibnamefont
  {Floyd}},\ }\emph {\bibinfo {title} {The Dynamics of Kerr Fields}},\
  \href@noop {} {Ph.D. thesis},\ \bibinfo  {school} {London Univ.}, \bibinfo
  {address} {London, England} (\bibinfo {year} {1973})\BibitemShut {NoStop}%
\bibitem [{\citenamefont {Penrose}(1973)}]{penrose1973}%
  \BibitemOpen
  \bibfield  {author} {\bibinfo {author} {\bibfnamefont {R.}~\bibnamefont
  {Penrose}},\ }\href@noop {} {\bibfield  {journal} {\bibinfo  {journal} {Ann.
  N. Y. Acad. Sci.}\ }\textbf {\bibinfo {volume} {224}},\ \bibinfo {pages}
  {125} (\bibinfo {year} {1973})}\BibitemShut {NoStop}%
\bibitem [{\citenamefont {Carter}(1968{\natexlab{a}})}]{carter1968}%
  \BibitemOpen
  \bibfield  {author} {\bibinfo {author} {\bibfnamefont {B.}~\bibnamefont
  {Carter}},\ }\href@noop {} {\bibfield  {journal} {\bibinfo  {journal} {Phys.
  Rev.}\ }\textbf {\bibinfo {volume} {174}},\ \bibinfo {pages} {1559} (\bibinfo
  {year} {1968}{\natexlab{a}})}\BibitemShut {NoStop}%
\bibitem [{\citenamefont {Mino}(2003)}]{mino2003}%
  \BibitemOpen
  \bibfield  {author} {\bibinfo {author} {\bibfnamefont {Y.}~\bibnamefont
  {Mino}},\ }\href@noop {} {\bibfield  {journal} {\bibinfo  {journal} {Phys.
  Rev. D}\ }\textbf {\bibinfo {volume} {67}},\ \bibinfo {pages} {084027}
  (\bibinfo {year} {2003})}\BibitemShut {NoStop}%
\bibitem [{\citenamefont {{Chandrasekhar}}(1983)}]{chandrasekharBH}%
  \BibitemOpen
  \bibfield  {author} {\bibinfo {author} {\bibfnamefont {S.}~\bibnamefont
  {{Chandrasekhar}}},\ }\href@noop {} {\emph {\bibinfo {title} {{The
  mathematical theory of black holes}}}}\ (\bibinfo  {publisher} {Clarendon
  Press},\ \bibinfo {year} {1983})\BibitemShut {NoStop}%
\bibitem [{\citenamefont {Schmidt}(2002)}]{schmidt2002}%
  \BibitemOpen
  \bibfield  {author} {\bibinfo {author} {\bibfnamefont {W.}~\bibnamefont
  {Schmidt}},\ }\href@noop {} {\bibfield  {journal} {\bibinfo  {journal}
  {Class. Quant. Grav.}\ }\textbf {\bibinfo {volume} {19}},\ \bibinfo {pages}
  {2743} (\bibinfo {year} {2002})}\BibitemShut {NoStop}%
\bibitem [{\citenamefont {Drasco}\ and\ \citenamefont
  {Hughes}(2004)}]{drasco2004}%
  \BibitemOpen
  \bibfield  {author} {\bibinfo {author} {\bibfnamefont {S.}~\bibnamefont
  {Drasco}}\ and\ \bibinfo {author} {\bibfnamefont {S.~A.}\ \bibnamefont
  {Hughes}},\ }\href@noop {} {\bibfield  {journal} {\bibinfo  {journal} {Phys.
  Rev. D}\ }\textbf {\bibinfo {volume} {69}},\ \bibinfo {pages} {044015}
  (\bibinfo {year} {2004})}\BibitemShut {NoStop}%
\bibitem [{\citenamefont {Fujita}\ and\ \citenamefont
  {Hikida}(2009)}]{fujita2009}%
  \BibitemOpen
  \bibfield  {author} {\bibinfo {author} {\bibfnamefont {R.}~\bibnamefont
  {Fujita}}\ and\ \bibinfo {author} {\bibfnamefont {W.}~\bibnamefont
  {Hikida}},\ }\href@noop {} {\bibfield  {journal} {\bibinfo  {journal} {Class.
  Quant. Grav.}\ }\textbf {\bibinfo {volume} {26}},\ \bibinfo {pages} {135002}
  (\bibinfo {year} {2009})}\BibitemShut {NoStop}%
\bibitem [{\citenamefont {Hackmann}\ \emph {et~al.}(2010)\citenamefont
  {Hackmann}, \citenamefont {L{\"a}mmerzahl}, \citenamefont {Kagramanova},\
  and\ \citenamefont {Kunz}}]{hackmann2010}%
  \BibitemOpen
  \bibfield  {author} {\bibinfo {author} {\bibfnamefont {E.}~\bibnamefont
  {Hackmann}}, \bibinfo {author} {\bibfnamefont {C.}~\bibnamefont
  {L{\"a}mmerzahl}}, \bibinfo {author} {\bibfnamefont {V.}~\bibnamefont
  {Kagramanova}}, \ and\ \bibinfo {author} {\bibfnamefont {J.}~\bibnamefont
  {Kunz}},\ }\href@noop {} {\bibfield  {journal} {\bibinfo  {journal} {Phys.
  Rev. D}\ }\textbf {\bibinfo {volume} {81}},\ \bibinfo {pages} {044020}
  (\bibinfo {year} {2010})}\BibitemShut {NoStop}%
\bibitem [{\citenamefont {Kagramanova}\ \emph {et~al.}(2010)\citenamefont
  {Kagramanova}, \citenamefont {Kunz}, \citenamefont {Hackmann},\ and\
  \citenamefont {L{\"a}mmerzahl}}]{kagramanova2010}%
  \BibitemOpen
  \bibfield  {author} {\bibinfo {author} {\bibfnamefont {V.}~\bibnamefont
  {Kagramanova}}, \bibinfo {author} {\bibfnamefont {J.}~\bibnamefont {Kunz}},
  \bibinfo {author} {\bibfnamefont {E.}~\bibnamefont {Hackmann}}, \ and\
  \bibinfo {author} {\bibfnamefont {C.}~\bibnamefont {L{\"a}mmerzahl}},\
  }\href@noop {} {\bibfield  {journal} {\bibinfo  {journal} {Phys. Rev. D}\
  }\textbf {\bibinfo {volume} {81}},\ \bibinfo {pages} {124044} (\bibinfo
  {year} {2010})}\BibitemShut {NoStop}%
\bibitem [{\citenamefont {Hackmann}\ and\ \citenamefont
  {L{\"a}mmerzahl}(2012)}]{hackmann2012}%
  \BibitemOpen
  \bibfield  {author} {\bibinfo {author} {\bibfnamefont {E.}~\bibnamefont
  {Hackmann}}\ and\ \bibinfo {author} {\bibfnamefont {C.}~\bibnamefont
  {L{\"a}mmerzahl}},\ }\href@noop {} {\bibfield  {journal} {\bibinfo  {journal}
  {Phys. Rev. D}\ }\textbf {\bibinfo {volume} {85}},\ \bibinfo {pages} {044049}
  (\bibinfo {year} {2012})}\BibitemShut {NoStop}%
\bibitem [{ker()}]{kerrgeo}%
  \BibitemOpen
  \href@noop {} {\enquote {\bibinfo {title} {{KerrGeodesics Mathematica
  package}},}\ }\bibinfo {howpublished}
  {(\href{http://bhptoolkit.org/KerrGeodesics/}{bhptoolkit.org/KerrGeodesics/})}\BibitemShut
  {NoStop}%
\bibitem [{\citenamefont {Mathisson}(1937)}]{mathisson1937}%
  \BibitemOpen
  \bibfield  {author} {\bibinfo {author} {\bibfnamefont {M.}~\bibnamefont
  {Mathisson}},\ }\href@noop {} {\bibfield  {journal} {\bibinfo  {journal}
  {Acta Phys. Polon.}\ }\textbf {\bibinfo {volume} {6}},\ \bibinfo {pages}
  {163} (\bibinfo {year} {1937})}\BibitemShut {NoStop}%
\bibitem [{\citenamefont {Papapetrou}(1951)}]{papapetrou1951}%
  \BibitemOpen
  \bibfield  {author} {\bibinfo {author} {\bibfnamefont {A.}~\bibnamefont
  {Papapetrou}},\ }\href {\doibase 10.1098/rspa.1951.0200} {\bibfield
  {journal} {\bibinfo  {journal} {Proc. Roy. Soc. Lond.}\ }\textbf {\bibinfo
  {volume} {A209}},\ \bibinfo {pages} {248} (\bibinfo {year}
  {1951})}\BibitemShut {NoStop}%
\bibitem [{\citenamefont {Dixon}(1964)}]{dixon1964}%
  \BibitemOpen
  \bibfield  {author} {\bibinfo {author} {\bibfnamefont {W.}~\bibnamefont
  {Dixon}},\ }\href {\doibase 10.1007/BF02734579} {\bibfield  {journal}
  {\bibinfo  {journal} {Il Nuovo Cimento (1955-1965)}\ }\textbf {\bibinfo
  {volume} {34}},\ \bibinfo {pages} {317} (\bibinfo {year} {1964})}\BibitemShut
  {NoStop}%
\bibitem [{\citenamefont {Tulczyjew}(1959)}]{tulczyjew1959}%
  \BibitemOpen
  \bibfield  {author} {\bibinfo {author} {\bibfnamefont {W.}~\bibnamefont
  {Tulczyjew}},\ }\href@noop {} {\bibfield  {journal} {\bibinfo  {journal}
  {Acta Phys. Pol.}\ }\textbf {\bibinfo {volume} {18}},\ \bibinfo {pages} {393}
  (\bibinfo {year} {1959})}\BibitemShut {NoStop}%
\bibitem [{\citenamefont {Dixon}(1970)}]{dixon1970}%
  \BibitemOpen
  \bibfield  {author} {\bibinfo {author} {\bibfnamefont {W.~G.}\ \bibnamefont
  {Dixon}},\ }\href {\doibase 10.1098/rspa.1970.0020} {\bibfield  {journal}
  {\bibinfo  {journal} {Proc. Roy. Soc. Lond.}\ }\textbf {\bibinfo {volume}
  {A314}},\ \bibinfo {pages} {499} (\bibinfo {year} {1970})}\BibitemShut
  {NoStop}%
\bibitem [{\citenamefont {Costa}\ and\ \citenamefont
  {Nat{\'a}rio}(2015)}]{costa2015}%
  \BibitemOpen
  \bibfield  {author} {\bibinfo {author} {\bibfnamefont {L.~F.~O.}\
  \bibnamefont {Costa}}\ and\ \bibinfo {author} {\bibfnamefont
  {J.}~\bibnamefont {Nat{\'a}rio}},\ }in\ \href {\doibase
  10.1007/978-3-319-18335-0_6} {\emph {\bibinfo {booktitle} {Equations of
  Motion in Relativistic Gravity}}}\ (\bibinfo  {publisher} {Springer},\
  \bibinfo {year} {2015})\ pp.\ \bibinfo {pages} {215--258},\ \Eprint
  {http://arxiv.org/abs/1410.6443} {arXiv:1410.6443 [gr-qc]} \BibitemShut
  {NoStop}%
\bibitem [{\citenamefont {Witzany}\ \emph {et~al.}(2019)\citenamefont
  {Witzany}, \citenamefont {Steinhoff},\ and\ \citenamefont
  {Lukes-Gerakopoulos}}]{spinpap}%
  \BibitemOpen
  \bibfield  {author} {\bibinfo {author} {\bibfnamefont {V.}~\bibnamefont
  {Witzany}}, \bibinfo {author} {\bibfnamefont {J.}~\bibnamefont {Steinhoff}},
  \ and\ \bibinfo {author} {\bibfnamefont {G.}~\bibnamefont
  {Lukes-Gerakopoulos}},\ }\href {\doibase 10.1088/1361-6382/ab002f} {\bibfield
   {journal} {\bibinfo  {journal} {Class. Quant. Grav.}\ }\textbf {\bibinfo
  {volume} {36}},\ \bibinfo {pages} {075003} (\bibinfo {year} {2019})},\
  \Eprint {http://arxiv.org/abs/1808.06582} {arXiv:1808.06582 [gr-qc]}
  \BibitemShut {NoStop}%
\bibitem [{\citenamefont {Ehlers}\ and\ \citenamefont
  {Rudolph}(1977)}]{ehlers1977}%
  \BibitemOpen
  \bibfield  {author} {\bibinfo {author} {\bibfnamefont {J.}~\bibnamefont
  {Ehlers}}\ and\ \bibinfo {author} {\bibfnamefont {E.}~\bibnamefont
  {Rudolph}},\ }\href {\doibase 10.1007/BF00763547} {\bibfield  {journal}
  {\bibinfo  {journal} {Gen. Relativ. Gravitation}\ }\textbf {\bibinfo {volume}
  {8}},\ \bibinfo {pages} {197} (\bibinfo {year} {1977})}\BibitemShut {NoStop}%
\bibitem [{\citenamefont {Marck}(1983)}]{marck}%
  \BibitemOpen
  \bibfield  {author} {\bibinfo {author} {\bibfnamefont {J.-A.}\ \bibnamefont
  {Marck}},\ }\href@noop {} {\bibfield  {journal} {\bibinfo  {journal} {Proc.
  R. Soc. Lond. A}\ }\textbf {\bibinfo {volume} {385}},\ \bibinfo {pages} {431}
  (\bibinfo {year} {1983})}\BibitemShut {NoStop}%
\bibitem [{\citenamefont {Tod}\ \emph {et~al.}(1976)\citenamefont {Tod},
  \citenamefont {de~Felice},\ and\ \citenamefont {Calvani}}]{tod1976}%
  \BibitemOpen
  \bibfield  {author} {\bibinfo {author} {\bibfnamefont {K.~P.}\ \bibnamefont
  {Tod}}, \bibinfo {author} {\bibfnamefont {F.}~\bibnamefont {de~Felice}}, \
  and\ \bibinfo {author} {\bibfnamefont {M.}~\bibnamefont {Calvani}},\ }\href
  {\doibase 10.1007/BF02728614} {\bibfield  {journal} {\bibinfo  {journal}
  {Nuovo Cim.}\ }\textbf {\bibinfo {volume} {B34}},\ \bibinfo {pages} {365}
  (\bibinfo {year} {1976})}\BibitemShut {NoStop}%
\bibitem [{\citenamefont {Hackmann}\ \emph {et~al.}(2014)\citenamefont
  {Hackmann}, \citenamefont {L{\"a}mmerzahl}, \citenamefont {Obukhov},
  \citenamefont {Puetzfeld},\ and\ \citenamefont {Schaffer}}]{hackmann2014}%
  \BibitemOpen
  \bibfield  {author} {\bibinfo {author} {\bibfnamefont {E.}~\bibnamefont
  {Hackmann}}, \bibinfo {author} {\bibfnamefont {C.}~\bibnamefont
  {L{\"a}mmerzahl}}, \bibinfo {author} {\bibfnamefont {Y.~N.}\ \bibnamefont
  {Obukhov}}, \bibinfo {author} {\bibfnamefont {D.}~\bibnamefont {Puetzfeld}},
  \ and\ \bibinfo {author} {\bibfnamefont {I.}~\bibnamefont {Schaffer}},\
  }\href@noop {} {\bibfield  {journal} {\bibinfo  {journal} {Phys. Rev. D}\
  }\textbf {\bibinfo {volume} {90}},\ \bibinfo {pages} {064035} (\bibinfo
  {year} {2014})}\BibitemShut {NoStop}%
\bibitem [{\citenamefont {Zelenka}\ \emph {et~al.}(2019)\citenamefont
  {Zelenka}, \citenamefont {Lukes-Gerakopoulos}, \citenamefont {Witzany},\ and\
  \citenamefont {Kopáček}}]{zelenka2019}%
  \BibitemOpen
  \bibfield  {author} {\bibinfo {author} {\bibfnamefont {O.}~\bibnamefont
  {Zelenka}}, \bibinfo {author} {\bibfnamefont {G.}~\bibnamefont
  {Lukes-Gerakopoulos}}, \bibinfo {author} {\bibfnamefont {V.}~\bibnamefont
  {Witzany}}, \ and\ \bibinfo {author} {\bibfnamefont {O.}~\bibnamefont
  {Kopáček}},\ }\href@noop {} {\  (\bibinfo {year} {2019})},\ \Eprint
  {http://arxiv.org/abs/1911.00414} {arXiv:1911.00414 [gr-qc]} \BibitemShut
  {NoStop}%
\bibitem [{\citenamefont {Grant}\ and\ \citenamefont
  {Flanagan}(2015)}]{grant2015}%
  \BibitemOpen
  \bibfield  {author} {\bibinfo {author} {\bibfnamefont {A.}~\bibnamefont
  {Grant}}\ and\ \bibinfo {author} {\bibfnamefont {{\'E}.~{\'E}.}\ \bibnamefont
  {Flanagan}},\ }\href@noop {} {\bibfield  {journal} {\bibinfo  {journal}
  {Class. Quant. Grav.}\ }\textbf {\bibinfo {volume} {32}},\ \bibinfo {pages}
  {157001} (\bibinfo {year} {2015})}\BibitemShut {NoStop}%
\bibitem [{\citenamefont {Witzany}(2017)}]{witzany2017}%
  \BibitemOpen
  \bibfield  {author} {\bibinfo {author} {\bibfnamefont {V.}~\bibnamefont
  {Witzany}},\ }\href {\doibase 10.1093/mnras/stx2520} {\bibfield  {journal}
  {\bibinfo  {journal} {Mon. Not. R. Astron. Soc.}\ }\textbf {\bibinfo {volume}
  {473}},\ \bibinfo {pages} {2434} (\bibinfo {year} {2017})},\ \Eprint
  {http://arxiv.org/abs/1709.03330} {arXiv:1709.03330 [gr-qc]} \BibitemShut
  {NoStop}%
\bibitem [{\citenamefont {Steinhoff}\ and\ \citenamefont
  {Puetzfeld}(2012)}]{steinhoff2012}%
  \BibitemOpen
  \bibfield  {author} {\bibinfo {author} {\bibfnamefont {J.}~\bibnamefont
  {Steinhoff}}\ and\ \bibinfo {author} {\bibfnamefont {D.}~\bibnamefont
  {Puetzfeld}},\ }\href {\doibase 10.1103/PhysRevD.86.044033} {\bibfield
  {journal} {\bibinfo  {journal} {Phys. Rev.}\ }\textbf {\bibinfo {volume}
  {D86}},\ \bibinfo {pages} {044033} (\bibinfo {year} {2012})},\ \Eprint
  {http://arxiv.org/abs/1205.3926} {arXiv:1205.3926 [gr-qc]} \BibitemShut
  {NoStop}%
\bibitem [{\citenamefont {Vines}\ \emph {et~al.}(2016)\citenamefont {Vines},
  \citenamefont {Kunst}, \citenamefont {Steinhoff},\ and\ \citenamefont
  {Hinderer}}]{vines2016}%
  \BibitemOpen
  \bibfield  {author} {\bibinfo {author} {\bibfnamefont {J.}~\bibnamefont
  {Vines}}, \bibinfo {author} {\bibfnamefont {D.}~\bibnamefont {Kunst}},
  \bibinfo {author} {\bibfnamefont {J.}~\bibnamefont {Steinhoff}}, \ and\
  \bibinfo {author} {\bibfnamefont {T.}~\bibnamefont {Hinderer}},\ }\href
  {\doibase 10.1103/PhysRevD.93.103008} {\bibfield  {journal} {\bibinfo
  {journal} {Phys. Rev.}\ }\textbf {\bibinfo {volume} {D93}},\ \bibinfo {pages}
  {103008} (\bibinfo {year} {2016})},\ \Eprint
  {http://arxiv.org/abs/1601.07529} {arXiv:1601.07529 [gr-qc]} \BibitemShut
  {NoStop}%
\bibitem [{\citenamefont {Hansen}(1974)}]{hansen1974}%
  \BibitemOpen
  \bibfield  {author} {\bibinfo {author} {\bibfnamefont {R.~O.}\ \bibnamefont
  {Hansen}},\ }\href@noop {} {\bibfield  {journal} {\bibinfo  {journal} {J.
  Math. Phys.}\ }\textbf {\bibinfo {volume} {15}},\ \bibinfo {pages} {46}
  (\bibinfo {year} {1974})}\BibitemShut {NoStop}%
\bibitem [{\citenamefont {Vines}(2018)}]{vines2018}%
  \BibitemOpen
  \bibfield  {author} {\bibinfo {author} {\bibfnamefont {J.}~\bibnamefont
  {Vines}},\ }\href@noop {} {\bibfield  {journal} {\bibinfo  {journal} {Class.
  Quant. Grav.}\ }\textbf {\bibinfo {volume} {35}},\ \bibinfo {pages} {084002}
  (\bibinfo {year} {2018})}\BibitemShut {NoStop}%
\bibitem [{\citenamefont {Carter}(1968{\natexlab{b}})}]{carter1968b}%
  \BibitemOpen
  \bibfield  {author} {\bibinfo {author} {\bibfnamefont {B.}~\bibnamefont
  {Carter}},\ }\href@noop {} {\bibfield  {journal} {\bibinfo  {journal}
  {Commun. Math. Phys.}\ }\textbf {\bibinfo {volume} {10}},\ \bibinfo {pages}
  {280} (\bibinfo {year} {1968}{\natexlab{b}})}\BibitemShut {NoStop}%
\bibitem [{\citenamefont {Pleba{\'n}ski}\ and\ \citenamefont
  {Demia{\'n}ski}(1976)}]{plebanski1976}%
  \BibitemOpen
  \bibfield  {author} {\bibinfo {author} {\bibfnamefont {J.~F.}\ \bibnamefont
  {Pleba{\'n}ski}}\ and\ \bibinfo {author} {\bibfnamefont {M.}~\bibnamefont
  {Demia{\'n}ski}},\ }\href@noop {} {\bibfield  {journal} {\bibinfo  {journal}
  {Ann. Phys.}\ }\textbf {\bibinfo {volume} {98}},\ \bibinfo {pages} {98}
  (\bibinfo {year} {1976})}\BibitemShut {NoStop}%
\bibitem [{\citenamefont {Connell}\ \emph {et~al.}(2008)\citenamefont
  {Connell}, \citenamefont {Frolov},\ and\ \citenamefont
  {Kubiz{\v{n}}{\'a}k}}]{connell2008}%
  \BibitemOpen
  \bibfield  {author} {\bibinfo {author} {\bibfnamefont {P.}~\bibnamefont
  {Connell}}, \bibinfo {author} {\bibfnamefont {V.~P.}\ \bibnamefont {Frolov}},
  \ and\ \bibinfo {author} {\bibfnamefont {D.}~\bibnamefont
  {Kubiz{\v{n}}{\'a}k}},\ }\href@noop {} {\bibfield  {journal} {\bibinfo
  {journal} {Phys. Rev. D}\ }\textbf {\bibinfo {volume} {78}},\ \bibinfo
  {pages} {024042} (\bibinfo {year} {2008})}\BibitemShut {NoStop}%
\bibitem [{\citenamefont {Van De~Meent}\ and\ \citenamefont
  {Warburton}(2018)}]{vandemeent2018}%
  \BibitemOpen
  \bibfield  {author} {\bibinfo {author} {\bibfnamefont {M.}~\bibnamefont {Van
  De~Meent}}\ and\ \bibinfo {author} {\bibfnamefont {N.}~\bibnamefont
  {Warburton}},\ }\href {\doibase 10.1088/1361-6382/aac8ce} {\bibfield
  {journal} {\bibinfo  {journal} {Class. Quant. Grav.}\ }\textbf {\bibinfo
  {volume} {35}},\ \bibinfo {pages} {144003} (\bibinfo {year} {2018})},\
  \Eprint {http://arxiv.org/abs/1802.05281} {arXiv:1802.05281 [gr-qc]}
  \BibitemShut {NoStop}%
\bibitem [{\citenamefont {Detweiler}\ and\ \citenamefont
  {Whiting}(2003)}]{detweiler2003}%
  \BibitemOpen
  \bibfield  {author} {\bibinfo {author} {\bibfnamefont {S.}~\bibnamefont
  {Detweiler}}\ and\ \bibinfo {author} {\bibfnamefont {B.~F.}\ \bibnamefont
  {Whiting}},\ }\href@noop {} {\bibfield  {journal} {\bibinfo  {journal} {Phys.
  Rev. D}\ }\textbf {\bibinfo {volume} {67}},\ \bibinfo {pages} {024025}
  (\bibinfo {year} {2003})}\BibitemShut {NoStop}%
\bibitem [{\citenamefont {Detweiler}(2008)}]{detweiler2008}%
  \BibitemOpen
  \bibfield  {author} {\bibinfo {author} {\bibfnamefont {S.}~\bibnamefont
  {Detweiler}},\ }\href@noop {} {\bibfield  {journal} {\bibinfo  {journal}
  {Phys. Rev. D}\ }\textbf {\bibinfo {volume} {77}},\ \bibinfo {pages} {124026}
  (\bibinfo {year} {2008})}\BibitemShut {NoStop}%
\bibitem [{\citenamefont {Sago}\ \emph {et~al.}(2008)\citenamefont {Sago},
  \citenamefont {Barack},\ and\ \citenamefont {Detweiler}}]{sago2008}%
  \BibitemOpen
  \bibfield  {author} {\bibinfo {author} {\bibfnamefont {N.}~\bibnamefont
  {Sago}}, \bibinfo {author} {\bibfnamefont {L.}~\bibnamefont {Barack}}, \ and\
  \bibinfo {author} {\bibfnamefont {S.}~\bibnamefont {Detweiler}},\ }\href@noop
  {} {\bibfield  {journal} {\bibinfo  {journal} {Phys. Rev. D}\ }\textbf
  {\bibinfo {volume} {78}},\ \bibinfo {pages} {124024} (\bibinfo {year}
  {2008})}\BibitemShut {NoStop}%
\bibitem [{\citenamefont {Mino}(2005)}]{mino2005}%
  \BibitemOpen
  \bibfield  {author} {\bibinfo {author} {\bibfnamefont {Y.}~\bibnamefont
  {Mino}},\ }\href@noop {} {\bibfield  {journal} {\bibinfo  {journal} {Progress
  of theoretical physics}\ }\textbf {\bibinfo {volume} {113}},\ \bibinfo
  {pages} {733} (\bibinfo {year} {2005})}\BibitemShut {NoStop}%
\bibitem [{\citenamefont {Hughes}\ \emph {et~al.}(2005)\citenamefont {Hughes},
  \citenamefont {Drasco}, \citenamefont {Flanagan},\ and\ \citenamefont
  {Franklin}}]{hughes2005}%
  \BibitemOpen
  \bibfield  {author} {\bibinfo {author} {\bibfnamefont {S.~A.}\ \bibnamefont
  {Hughes}}, \bibinfo {author} {\bibfnamefont {S.}~\bibnamefont {Drasco}},
  \bibinfo {author} {\bibfnamefont {E.~E.}\ \bibnamefont {Flanagan}}, \ and\
  \bibinfo {author} {\bibfnamefont {J.}~\bibnamefont {Franklin}},\ }\href@noop
  {} {\bibfield  {journal} {\bibinfo  {journal} {Phys. Rev. Lett.}\ }\textbf
  {\bibinfo {volume} {94}},\ \bibinfo {pages} {221101} (\bibinfo {year}
  {2005})}\BibitemShut {NoStop}%
\bibitem [{\citenamefont {Sago}\ \emph {et~al.}(2006)\citenamefont {Sago},
  \citenamefont {Tanaka}, \citenamefont {Hikida}, \citenamefont {Ganz},\ and\
  \citenamefont {Nakano}}]{sago2006}%
  \BibitemOpen
  \bibfield  {author} {\bibinfo {author} {\bibfnamefont {N.}~\bibnamefont
  {Sago}}, \bibinfo {author} {\bibfnamefont {T.}~\bibnamefont {Tanaka}},
  \bibinfo {author} {\bibfnamefont {W.}~\bibnamefont {Hikida}}, \bibinfo
  {author} {\bibfnamefont {K.}~\bibnamefont {Ganz}}, \ and\ \bibinfo {author}
  {\bibfnamefont {H.}~\bibnamefont {Nakano}},\ }\href@noop {} {\bibfield
  {journal} {\bibinfo  {journal} {Progress of theoretical physics}\ }\textbf
  {\bibinfo {volume} {115}},\ \bibinfo {pages} {873} (\bibinfo {year}
  {2006})}\BibitemShut {NoStop}%
\end{thebibliography}%

\appendix

\section{Angular variables and averaging} \label{app}

\subsection{Coordinate transformation}
Start by re-expressing $\di r /\di \lambda, \di \vartheta/\di \lambda$ as
\begin{align}
     & \frac{\di r}{\di \lambda} = \pm Y_r \sqrt{(r-r_{1\rm g} - \delta r_{1})(r_{2\rm g} + \delta r_{2}-r)} \,,
     \\
     & \frac{\di \vartheta}{\di \lambda} = \mp \frac{Y_\vartheta}{\sin \vartheta} \sqrt{(\zeta - \zeta_1)(\zeta_2 - \zeta)} \,,
     \\
     & Y_r(r,\vartheta,\tilde{s}^{CD}) \equiv \sqrt{\frac{{w_r'}^2 - e_{0r} e_{C;r}^\kappa e_{D \kappa} \tilde{s}^{CD}}{(r-r_{1\rm g} - \delta r_{1})(r_{2\rm g} + \delta r_{2}-r)}}\,,
     \\
     & Y_\vartheta(r,\vartheta,\tilde{s}^{CD}) \equiv -\sin \vartheta \sqrt{\frac{{w_\vartheta'}^2 - e_{0\vartheta} e_{C;\vartheta}^\kappa e_{D \kappa} \tilde{s}^{CD}}{(\zeta - \zeta_1)(\zeta_2 - \zeta)}}\,,\\
     & \zeta \equiv \cos \vth,\, \zeta_i \equiv \cos(\vth_{i\rm g} + \delta \vth_i),\, i=1,2\,.
\end{align}
The expressions $Y_r,Y_\vartheta$ expanded to $\mathcal{O}(s)$ are now regular and nonzero for the entire trajectory. However, they will generally have an $\mathcal{O}(s^2)$ term that diverges as $1/\sqrt{y-y_{\rm t}}$ around the turning point. I assume that one can always introduce an $\mathcal{O}(s)$ shift to the background congruence constants $K_{\rm c},L_{\rm c}, E_{\rm c}$ so as to cancel this term. In practice, I simply expand to linear order in $s$ as $Y_y = Y_{y0} + \delta Y_y + \mathcal{O}(s^2)$ and discard higher-order terms. The $Y_{y0}$ are most practically expressed as
\begin{align}
    &Y_{r0} = \sqrt{(1-E^2)(r_0-r_{3\rm g})(r_0-r_{4\rm g})} \,,\\
    &Y_{\vartheta 0} = \sqrt{a^2 (1 -E^2)(z_{+\rm g} -\zeta_0^2)}\,,
\end{align}
where $\zeta_0\equiv \cos \vth_0$ (compare with section \ref{sec:kerrgeo}). The spin-induced corrections $\delta Y_y$ can then be written as
\begin{subequations}
\begin{align}
    &\delta Y_r = \frac{\mathcal{K}_r - \Delta \mathcal{J}_r }{2 Y_{r0} (r_{2\rm g} - r_0)(r_0 - r_{1\rm g})} + \frac{\partial Y_{r0}}{\partial r_0} \delta r \,,
    \\ 
    &\delta Y_\vartheta = \frac{\mathcal{K}_\vth - (1-\zeta_0^2) \mathcal{J}_\vth }{2 Y_{\vartheta 0} (z_{-\rm g} - \zeta^2_0)} + \frac{\partial Y_{\vartheta0}}{\partial \vartheta_0} \delta \vartheta\,,
    \\
    & \mathcal{J}_r \equiv \frac{2s_\parallel \mathcal{G} + \Delta (K + r^2) e_{0 r} {e}_{C;r}^\kappa  {e}_{D \kappa} \tilde{s}^{CD}}{K + r^2}\Big|_{\substack{r=r_0\\ \vartheta=\vartheta_0}}
    \,,\\
    & \mathcal{J}_\vartheta \equiv \frac{2s_\parallel\mathcal{H} + e_{0 \vartheta} {e}_{C;\vartheta}^\kappa  {e}_{D \kappa} \tilde{s}^{CD}}{K -   a^2 \cos^2 \!\vartheta}\Big|_{\substack{r=r_0\\ \vartheta=\vartheta_0}}
    \,,
    \\ 
    &\mathcal{K}_r \equiv \left[\delta r_1 (r_{2\rm g} - r_0) - \delta r_2 (r_0 - r_{1\rm g})\right]Y_{r0}^2\,,
    \\ 
    & \mathcal{K}_\vth \equiv \left[\delta \zeta_1(\zeta_0 - \sqrt{z_{- \rm g}}) + \delta \zeta_2(\zeta_0 + \sqrt{z_{- \rm g}})\right] Y_{\vartheta 0}^2\,,
    \\ 
    &\frac{\partial Y_{r0}}{\partial r_0} = \frac{\sqrt{1 - E^2/ } (2 r_0 - r_{3\rm g} - r_{4\rm g}) }{\sqrt{(r_0-r_{3\rm g})(r_0-r_{4\rm g})}} \,,
    \\
    &\frac{\partial Y_{\vartheta0}}{\partial \vartheta_0} = \frac{ \sqrt{a^2(1 - E^2/ )(1-\zeta_0)} \zeta_0}{\sqrt{(z_{+\rm g}-\zeta_0^2)}} \,,
\end{align}
\end{subequations}
where $\mathcal{G},\mathcal{H}$ were defined in \eqref{eq:turnshifts}. Note that I am using the same fiducial geodesic $K = K_{\rm so} + 2   a s_\parallel {\rm sgn}(L_{\rm so}- a E_{\rm so}), E = E_{\rm so}, L = L_{\rm so}$ as in the computation of the turning-point shifts. Additionally, one must sew the functions $\mathcal{J}_y(\chi_r,\chi_\vartheta)$ from parts where the congruence four-velocity $u^\mu_{\rm c} = e^\mu_{0}$ always has the same signature as the actual four-velocity $u^\mu$.

The expressions for $\delta Y_y$ avoid $1/\cos(\chi_y)^2$ singularities by the numerators of the first terms vanishing at $\chi_y = \uppi/2,3\uppi/2$; this is easily seen by comparing with the turning-point formulae \eqref{eq:deltart} and \eqref{eq:deltatht}. Nevertheless, we must still verify that the numerator of the first term in $\delta Y_y$ has a zero derivative with respect to $\chi_y$ at $\chi_y = \uppi/2,3\uppi/2$, otherwise a $1/\cos(\chi_y)$ divergence occurs. It turns out that the derivatives do not vanish only when $K_{\rm c},E_{\rm c},L_{\rm c} = K,E,L$. I thus choose the congruence constants so that tetrad turning points occur outside of the envelope of the spin-perturbed motion (note that this also requires inserting an accordingly shifted connection into the turning-point formulae \eqref{eq:turnshifts} so that $\mathcal{X}^{(y)}_{\kappa C} e_{D}^\kappa\to e_{0y} e_{C \kappa;y}e_{D}^\kappa|_{y=y_{\rm gt}}$).

Let us now use the transformation \eqref{eq:angparam} to express $\di y/\di \lambda$ in terms of $\di \chi_r/\di \lambda$, $\di \chi_\vartheta/\di \lambda$, and $\di \phi/\di \lambda$, and we finally obtain the functions $f_y,\delta f_y$ so that $\di \chi_y/\di \lambda = f_y + \delta f_y$
\begin{align}
     &f_y(\chi_y) = Y_{y0}\,,\\
     &\delta f_y = \delta Y_y -  \frac{2 \sigma_y}{(y_{1\rm g} - y_{2\rm g})\cos \chi_y}  \,, \\
     &\sigma_y \equiv (h_r + h_\vartheta) \frac{\partial \delta y}{\partial \phi} + Y_{x0} \frac{\partial \delta y}{ \partial \chi_x}\,,
\end{align}
where $x,y = r,\vartheta$ or $\vartheta,r$. 

Now we see that the second term in $\delta f_y$ is singular because the turning points are not separable; furthermore, it can be shown that no global choice of $\delta y$ can transform these singularities away without introducing stronger ones. However, these terms can be eliminated by holding the expressions for $\di \chi_y/ \di \lambda$ non-expanded in $s$. In return, one can then see that the $\mathcal{O}(s)$ singular terms in $\delta f_y$ correspond to the regular non-expanded $\di \chi_y/ \di \lambda$ receiving a $\mathcal{O}(1)$ spin-correction within an $\mathcal{O}(s)$ interval near the turning points. Nevertheless, I find it simpler to keep the formulae in expanded form since these singular terms end up averaging out to zero in any expression of interest (see section \ref{sec:singavg}). 

\subsection{Homogeneous angles and geodesic averaging} \label{app:avg}
Let us start with the $\tilde{s}^{CD}=0$ problem. I define the homogeneous angle coordinates
\begin{align}
    &\psi_r(\chi_r) \equiv \Upsilon_{r \rm g}\int_0^{\chi_r} \frac{\di \chi'_r}{f_r(\chi'_r)}\,,\\
    &\psi_\vartheta(\chi_\vartheta) \equiv \Upsilon_{\vartheta \rm g}\int_0^{\chi_\vartheta} \frac{\di \chi'_\vartheta}{f_\vartheta(\chi'_\vartheta)}\,,\\
    \begin{split}
    & \psi_\phi(\phi,\chi_r,\chi_\vartheta) \equiv \phi + \int_0^{\chi_r} \frac{\langle h_r\rangle - h_r(\chi'_r)}{f_r} \di \chi'_r 
    \\
    & \quad\quad\quad\quad\quad\quad\;\;\;+ \int_0^{\chi_\vartheta} \frac{\langle h_\vartheta \rangle - h_\vartheta(\chi'_\vartheta)}{f_\vartheta} \di \chi'_\vartheta\,, \end{split}
    \\
    & \langle h_y \rangle \equiv \frac{\Upsilon_{y \rm g}}{2 \uppi}\int_0^{2\uppi} \frac{h_y}{f_y} \di \chi_y \,, \\
    & \Upsilon_{y \rm g} \equiv 2\uppi \left(\int_0^{2\uppi} \frac{\di \chi_y}{f_y} \right)^{-1}\,, \\
\end{align}
Consequently, the equations of motion for the angles $\vec{\psi} = (\psi_r,\psi_\vartheta,\psi_\varphi)$ read
\begin{align}
    & \frac{\di \vec{\psi}}{\di \lambda} = \vec{\Upsilon} \,,
\end{align}
where $\vec{\Upsilon} = (\Upsilon_{r \rm g}, \Upsilon_{\vartheta \rm g},\Upsilon_{\phi \rm g}),\, \Upsilon_{\phi \rm g} \equiv \langle h_r \rangle + \langle h_\vartheta \rangle$.

If for every $\vec{k} \in \mathbb{Z}^3$ we have $\vec{k} \cdot \vec{\Upsilon} \neq 0 $ (the motion is not resonant), then the long-term average of any function $j(\psi_r,\psi_\vartheta,\psi_\varphi)$ over the trajectory can be written as an average over angles
\begin{align}
\begin{split}
    \langle j(\vec{\psi}(\lambda)) \rangle_{\rm g} 
    & \equiv \lim_{\Lambda \to \infty} \frac{1}{ \Lambda} \int_{0}^\Lambda j(\vec{\psi}(\lambda)) \di \lambda
    \\
    & = \frac{1}{(2 \uppi)^3}\int_{(0,2\uppi]^3} j(\vec{\psi}) \di^3 \psi
    \\
    & = \frac{\Upsilon_{r \rm g} \Upsilon_{\vartheta \rm g}}{(2 \uppi)^3}\int_{(0,2\uppi]^3} \frac{j(\chi_r,\chi_\vartheta,\phi)}{f_r f_\vartheta} \di \chi_r \di \chi_\vartheta \di \phi \,, 
\end{split} \label{eq:avg}
\end{align}
where in the last equality I have used the Change of variables theorem. One of the consequences of the formula above is the fact that the only components of spin that end up having any influence in long-term geodesic averages of quantities linear in spin are $\tilde{s}^{12} = -\tilde{s}^{21} = s_\parallel$.

\subsection{Fundamental frequencies}
Now let us consider the equations of motion under the spin perturbation, in the homogeneous angle variables we obtain
\begin{subequations}
\begin{align}
    &\frac{\di \psi_r}{\di \lambda} = \Upsilon_{r \rm g}\left( 1 +  \frac{\delta f_r}{f_r}\right)\,,\\
    &\frac{\di \psi_\vartheta}{\di \lambda} = \Upsilon_{\vartheta \rm g}\left( 1 +  \frac{\delta f_\vartheta}{f_\vartheta}\right) \,,\\
    &\frac{\di \psi_\phi}{\di \lambda} = \Upsilon_{\phi \rm g} \,.
\end{align} \label{eq:geo-angle}
\end{subequations}
This system of equations can be put back into homogeneous form by a near-identity transform $\vec{\Psi} = \vec{\psi} + \vec{\xi}$, where the formal solution for $\vec{\xi}$ is given as \citep[see, e.g.][]{arnold2007}
\begin{subequations} \label{eq:xisol}
\begin{align}
    &\xi_r = \sum_{\vec{k}\neq 0} \frac{1}{i \vec{k} \cdot \vec{\Upsilon}}\mathcal{F}^{\vec{k}}\left[\frac{\delta f_r}{f_r}\right] \exp(i \vec{k} \cdot \vec{\psi})\,,\\
    &\xi_\vartheta = \sum_{\vec{k}\neq 0} \frac{1}{i \vec{k} \cdot \vec{\Upsilon}}\mathcal{F}^{\vec{k}}\left[\frac{\delta f_\vartheta}{f_\vartheta}\right] \exp(i \vec{k} \cdot \vec{\psi})\,,\\
    &\xi_\phi = 0 \,,\\
    &\mathcal{F}^{\vec{k}}[j] \equiv \frac{1}{(2 \uppi)^3} \int_{(0,2\uppi]^3} j(\vec{\psi}) \exp(-i\vec{k}\cdot \vec{\psi}) \di^3 \psi\,.
\end{align}
\end{subequations}
This solution does not exist whenever the perturbation has a non-zero Fourier coefficient for a $\vec{k}$ such that $\vec{k}\cdot \vec{\Upsilon} =0$. Assuming for now that we are not dealing with such resonant cases, the new angle variables will fulfill the equations of motion
\begin{align}
    &\frac{\di \Psi_r}{\di \lambda} = \Upsilon_{r \rm g}\left( 1 +  \left\langle\frac{\delta f_r}{f_r}\right\rangle_{\rm g}\right)\,,\\
    &\frac{\di \Psi_r}{\di \lambda} = \Upsilon_{\vartheta \rm g}\left( 1 +  \left\langle\frac{\delta f_\vartheta}{f_\vartheta}\right\rangle_{\rm g} \right) \,.
\end{align}
In other words, by using equation \eqref{eq:avg} it is possible to compute the shift to the fundamental frequencies without the knowledge of a closed-form transformation to either of the angle coordinates $\vec{\psi}$ or $\vec{\Psi}$.

\subsection{Spin-perturbed averaging}\label{app:spavg}
Let us now briefly derive the general formula for averaging of separable functions over the spin-perturbed trajectory. Assume that we want to compute the average of a quantity $n(y)$, where again $y$ is either $r$ or $\vartheta$. We compute up to $\mathcal{O}( s^2)$
\begin{align}
\begin{split}
    \langle n(y) \rangle_{\rm sp}  
    & = \frac{1}{(2 \uppi)^3} \int n(y(\vec{\Psi})) \di^3 \Psi 
    \\
    & = \frac{1}{(2 \uppi)^3} \int n(y_0) \di^3 \Psi + \frac{1}{(2 \uppi)^3} \int n'(y_0) \delta y \di^3 \Psi\,.
\end{split}
\end{align}
Now, let us write $\tilde{n}(\psi_y) = n(y_0(\chi_y(\psi_y)))$ and further re-express
\begin{align}
\begin{split}
    & \int \tilde{n}(\psi_y) \di^3 \Psi = \int \tilde{n}(\Psi_y - \xi_y) \di^3 \Psi 
      \\
      &= \int \tilde{n}(\Psi_y) \di^3 \Psi - \int \tilde{n}'(\Psi_y) \xi_y \di^3 \Psi 
    \\ 
    & = \int \tilde{n}(\Psi_y) \di^3 \Psi + \int \tilde{n} \frac{\partial \xi_y}{\partial \Psi_y} \di^3 \Psi 
    \\
    &= \int \tilde{n}(\Psi_y) \di^3 \Psi + \int \left( \left\langle \frac{\delta f_y}{f_y} \right\rangle_{\rm g} - \frac{\delta f_y}{f_y} \right) \tilde{n}(\Psi_y) \di^3 \Psi\,,
\end{split}
\end{align}
where I have disposed of various boundary terms that vanish due to the periodicity of the involved functions and further used the property
\begin{align}
    \frac{\partial \xi_y}{\partial \Psi_r} \Upsilon_{r \rm g} + \frac{\partial \xi_y}{\partial \Psi_\vartheta} \Upsilon_{\vartheta \rm g} =  \left(\left\langle \frac{\delta f_y}{f_y} \right\rangle_{\rm g} - \frac{\delta f_y}{f_y} \right) \Upsilon_{y \rm g}\,.
\end{align}
One last point to realize is the fact that averaging an $\mathcal{O}(s)$ term over the spin-perturbed trajectory can be replaced by geodesic averages up to $\mathcal{O}( s^2)$. Consequently, all of the terms in $\langle n(y) \rangle_{\rm sp} $ are now expressible as closed-form averages over the geodesics, which can be summarized as
\begin{align}
\begin{split}
    \langle n(y) \rangle_{\rm sp}  =& \left(1+\left\langle \frac{\delta f_y}{f_y} \right\rangle_{\rm \!\! g} \right)\langle n(y_0) \rangle_{\rm g} + \langle n'(y_0) \delta y \rangle_{\rm g} \\
    &- \left\langle n(y_0) \frac{\delta f_y}{f_y} \right\rangle_{\rm \!\! g}\,.
\end{split}
\end{align}
It should be noted that for a non-separable function of both $r$ and $\vartheta$ additional $\xi$-dependent terms would emerge in the average.

\subsection{Averaging singular terms} \label{sec:singavg}
I showed that the shifts to fundamental frequencies are extracted from the system by computing geodesic averages. However, I also have to show that the singularities in $\delta f_y$ do not spoil the finiteness and non-ambiguity of the averages. 

All of the averages with $\delta f_y$ we need to compute are of the type $\langle j(\chi_y) \delta f_y \rangle_{\rm g}$. We write 
\begin{align}
\begin{split}
    &\langle j(\chi_y) \delta f_y \rangle_{\rm g} 
     \\ &\equiv  \frac{\Upsilon_{r \rm g} \Upsilon_{\vartheta \rm g}}{(2 \uppi)^3}\int \frac{j(\chi_r) }{f_r f_\vartheta} \frac{-2 \sigma_y}{\cos{\chi_y}(y_{1\rm g} - y_{2 \rm g})} \di \chi_r \di \chi_\vartheta \di \phi 
     \\& \quad \;+ \langle j\, \delta Y_y \rangle_{\rm g}
    \\
     &= \int k(\chi_r,\chi_\vartheta) \left( \int_0^{2\uppi} \frac{\partial \delta y}{\partial \phi} \di \phi \right) \di \chi_r \di \chi_\vartheta 
     \\
     & \quad \;+ \int l(\chi_y) \left( \int_0^{2\uppi} \frac{\partial \delta y}{\partial \chi_x} \di \chi_x \right) \di \chi_y \di \phi
     \\
     & \quad \; + \langle j\, \delta Y_y \rangle_{\rm g}
    \\
     & =  \langle j\, \delta Y_y \rangle_{\rm g}\,,
\end{split}
\end{align}
where $k,l$ are some functions of their variables. In other words, one can ignore the $\sim \partial \delta y/\partial \phi, \partial \delta y/\partial \chi_x$ terms in the average. 

If we want to construct the vector $\vec{\xi}$, we need to find all the Fourier coefficients of $\delta f_y/f_y$. It turns out that these will all be finite and uniquely defined as Cauchy principal-value integrals. Nevertheless, there still might be issues with the convergence of the sums given in \eqref{eq:xisol}. In that case, it is possible to instead hold $\di \chi_y/\di\lambda$ in non-expanded form and compute the Fourier coefficients of the regular expression $(\di \chi_y/\di\lambda)/f_y - 1$.

\end{document}